# Ultrafast Inter-Aggregate Energy Transfer Drives Efficient Charge Generation in Organic Homojunctions

*Shanchao OUYANG[1,2], Hang Yuen HO[1,2], Sudhi MAHADEVAN[1,2], Ziyao YUE[1,2], Haochen LIU[1], Zhiqiang GUAN[2,3], Chun-Sing LEE[2,3], Shu Kong SO[4] and Sai-Wing TSANG[1,2,5*]*

1 Department of Materials Science and Engineering, City University of Hong Kong, Hong Kong SAR, PR China

2 Centre of Super-Diamond and Advanced Films, City University of Hong Kong, Hong Kong SAR, PR China.

3 Department of Chemistry, City University of Hong Kong, Hong Kong SAR, PR China

4 Department of Physics and Institute of Advanced Materials, Hong Kong Baptist University, Kowloon Tong, Hong Kong SAR, PR China

5 The City University of Hong Kong Shenzhen Research Institute, Shenzhen, PR China

Email:saitsang@cityu.edu.hk

## Abstract

Y6-based homojunction organic photovoltaic devices have recently achieved exceptional power conversion efficiencies, surpassing those of conventional organic photovoltaic materials. Although unusually long exciton diffusion lengths and efficient exciton dissociation have been reported, the fundamental mechanism underlying this outstanding performance remains under debate. Here, we report that structurally distinct aggregates in Y6 thin films exhibit complementary charge-transfer (CT) characters that enable directional energy transfer and enhanced exciton delocalization. Using ultrafast spectroscopy, we identify highly efficient Förster resonance energy transfer (FRET) between Y6 aggregates, with an ensemble-averaged efficiency of 65% occurring on a sub-picosecond timescale (<1 ps). This aggregate-mediated energy-transfer pathway drives long-range energy migration in Y6 thin films and accounts for the remarkable external quantum efficiencies observed in homojunction devices, even in the absence of conventional donor-acceptor bulk heterojunction morphologies and despite substantial fractions of weak-CT components. Here, using Y6 as a model non-fullerene acceptor (NFA) system, we show that structurally distinct aggregates coexisting in NFA thin films possess complementary CT characters, which are expected to enable directional energy transfer and enhanced exciton delocalization. We establish a general multi-component FRET framework that quantitatively links molecular aggregation, CT character, and excited-state dynamics in systems with multiple packing motifs. This work provides insight into the development of next-generation high-performance organic electronic materials and devices.

# 1. Introduction

Small-molecule and polymeric organic semiconductors have emerged as a versatile platform for optoelectronic technologies, owing to their synthetically tunable electronic structures, solution processability, and intrinsic mechanical flexibility. The continuous advances in molecule design, heterojunction architecture, and device engineering have driven remarkable performance improvement in various organic optoelectronic devices [1-7]. In particular, recent development of non-fullerene acceptors (NFAs) has reshaped the organic solar cell (OSC) landscape. Overcoming the intrinsic limitations of fullerene-based acceptors. NFAs enable more precise control over energy level, absorption spectra, and intermolecular interactions, improving donor-acceptor complementarity and enabling more efficient photon-to-charge conversion [8]. Early development of NFAs centered around an A-D-A molecular architecture exemplified by ITIC, first reported by Zhan et al. in 2015 [9]. Compared to fullerene-based acceptors, ITIC acceptors offer stronger band-edge optical absorption and reduce the heterojunction energy offset, thereby dramatically improving OSC performance, with PCEs rising from 6.8% to approximately 13% [10]. A subsequent breakthrough was achieved with the development of Y-series acceptors featuring an A-DA'D-A structure, highlighted by the Y6 molecule, whose device PCE reached nearly 15% in 2019 [11]. More recently, Y6 derivative-based OSCs have set new records with efficiencies above 20% [12-14]. Moving forward, it is crucial not only to optimize the device architectures incorporating Y6 and its derivatives but also to develop theoretical frameworks that elucidate the fundamental mechanisms underlying their exceptional performance.

Highly efficient OSCs predominantly employ bulk heterojunction (BHJ) architectures, in which donor and acceptor materials are intimately interpenetrated to form extensive donor/acceptor (D/A) interfaces at the nanoscale [15]. Together with the energy level offset at the D/A interface, BHJ revolutionizes OSCs by overcoming the strong exciton binding energy typically observed in organic semiconductors [16]. Strikingly, it has recently been reported that homojunction devices containing only Y6 acceptors have achieved PCE over 4% and external quantum efficiencies (EQE) exceeding 20%, more than one order of magnitude higher than those previously reported for homojunction organic devices [17]. In addition, heterojunction devices with a low donor ratio (around 10 wt%) have even demonstrated PCE above 10% with EQE approaching 60% [18,19]. In these devices, the active layer evidently does not form a nanoscale BHJ structure, and the donor is solely responsible for improving charge-collection efficiency. Such exceptional photovoltaic performance has been attributed to the strong CT character of a particular Y6 dimer, as demonstrated theoretically [20]. However, it is worth noting that about 30% of the Y6 molecules do not aggregate but remain isolated in a solid-state homojunction thin film [20,21]. Compared to aggregated Y6 dimers, these isolated Y6 molecules exhibit weaker CT characteristics [22-24], which are ostensibly unlikely to contribute to exciton dissociation and positively to device performance.

The relatively high EQE observed in Y6 homojunction devices suggests that photon energy is efficiently transferred from weak to strong CT species within the active layer. Therefore, it is crucial to investigate and thoroughly understand the energy transfer mechanisms within the active layer. In general, energy-transfer mechanisms can be broadly categorized as radiative photon recycling or non-radiative processes, such as Dexter energy transfer and Förster resonance energy transfer (FRET). Due to the low radiative efficiency and large Stokes shifts in organics, they are intrinsically unfavorable for photon recycling [25], which has rarely been invoked in OSCs. Whereas Dexter charge transfer relies on wavefunction overlap between molecules, resulting in an exponential decay of transfer efficiency with increasing intermolecular distance, thereby limiting it to a short-range process operating only between neighboring molecules [26,27]. On the other hand, FRET is mediated by molecular dipole-dipole (Coulombic) interactions. It does not require physical contact between molecules but is highly sensitive to the spectral overlap between UV-Vis absorption and photoluminescence (PL). Accordingly, previously reported FRET processes in OSCs have primarily focused on energy transfer between organic donors and acceptors with distinct optical absorption spectra [28-36]. In addition, self-FRET, namely energy transfer between identical molecules, has been proposed to account for the observed increase in the exciton diffusion length and improved photovoltaic performance in bilayer OSCs [37,38]. Nevertheless, similar to photon recycling, Self-FRET is rare in organic semiconductors because of the large separation between their absorption and emission spectra.

In this study, we elucidate how distinct aggregate species mediate energy transfer and charge generation in Y6 homojunction thin films. To systematically probe this process, we controlled the aggregation of Y6 molecules by fabricating Y6:polymer solid-solution thin films with varying Y6 loading ratios. The aggregation behavior in these films was characterized by UV-Vis absorption and PL spectroscopy, combined with Franck-Condon weighted density (FCWD) analysis. Exciton recombination kinetics and exciton dynamics were probed by time-resolved PL (TRPL) and transient absorption spectroscopy (TAS), respectively. Theoretical simulations were used to determine the CT character of the species participating in the energy-transfer process. Based on these experimental and theoretical results, we reveal an aggregate-mediated efficient FRET process in Y6 homojunction thin films. Our multi-component FRET model demonstrates that FRET efficiency is governed not only by donor-acceptor excitonic coupling, but strongly by species weight densities. These findings provide new mechanistic insights into the design of future molecular architectures for organic electronic devices.

## 2. Results

### 2.1 Clues from aggregation progression

**Figure 1(a)** shows the molecular structures of materials used in this study. Polyvinylcarbazole (PVK), a wide-bandgap polymer with strong miscibility with Y6 [39], was blended with Y6 at different loading ratios in solution and spin-coated to fabricate solid-solution thin films. As illustrated in **Figure 1(b)**, increasing the Y6 loading in the PVK matrix reduces the average intermolecular distance between Y6 molecules, promoting the evolution from isolated molecules to aggregates with distinct configurations.

**Figures 1(c)** and **1(d)** show the corresponding UV-Vis absorption and PL spectra of films at various wt% of Y6 in PVK. For comparison, the absorption spectrum of Y6 in solution is also shown in **Figure 1(c)** to illustrate the dilution effect imparted by the PVK matrix. Notably, the absorption and emission edges exhibit contrasting shifts between films with low (≤10 wt%) and high (>10 wt%) Y6 loadings, as indicated by the arrows. At loadings less than 10 wt%, the absorption spectra features resemble well with that in the solution. This indicates effective molecular isolation within the PVK matrix, suggesting that aggregation remains limited and molecules predominantly exist as isolated entities. Above 10 wt%, pronounced red shifts and broadening of the absorption spectra emerge, with the most striking spectral evolution occurring between 10 wt% and 20 wt%. Further concentration increments beyond 20 wt% induce additional broadening, albeit less dramatically. In contrast, as shown in **Figure 1(d)**, a pronounced PL emission edge shift is observed at low Y6 loading ratios from 1 wt% to 10 wt%, and the shift becomes subtle with further increasing the loadings. The changes in absorption and PL spectra with respect to the Y6 loadings are summarized in **Figure 1(e)**.

The evolution of the absorption and PL spectra with increasing Y6 loading provides valuable insight into the correlation between molecular packing and energy transfer in Y6 thin films. First, although the average intermolecular distance decreases as the Y6 loading increases from 1 wt% to 10 wt%, most Y6 molecules remain largely isolated. Meanwhile, the pronounced PL redshift in this regime suggests the emergence of a distance sensitive long-range energy-transfer process, through which excitons can migrate to lower-energy states on neighboring isolated molecules [40]. This process is likely faster than radiative recombination from the initially excited molecules. In addition, the abrupt redshift and broadening of the absorption spectra from 10 wt% to 20 wt% indicate the existence of a threshold intermolecular distance for molecular aggregation. While intermolecular distance is smaller than this threshold, aggregates form abruptly from previously isolated molecules; thereafter, the aggregate population increases progressively with Y6 loading, as evidenced by continued spectral broadening and persistent absorption redshifts. Moreover, the broadened absorption features and sustained redshift suggest not only an increasing aggregate population, but also the emergence of diverse dipole-dipole alignments enabled by the distinctive architecture of Y6 aggregates. It has been reported previously that different configurations of Y6 dimers exhibit significantly different CT characters [20-24]. Therefore, deconvoluting the contributions of distinct aggregate packing arrangements to the overall energy-transfer process is imperative for understanding energy migration within the Y6 system. Accordingly, our mechanistic investigation focuses on two key questions: (i) how energy-transfer pathways evolve during aggregation and (ii) how specific aggregate configurations contribute to these pathways.

## 2.2 The energy transfer mechanisms between Y6 aggregates

To determine the operative energy-transfer pathways, we first investigated exciton recombination characteristics across different degrees of aggregation in solid-solution films. As shown in **Eq. 1** and **Eq. 2**, the radiative ($K_r$) and non-radiative ($K_{nr}$) decay rates can be derived from PL lifetimes ($\tau$) and PL quantum yields (PLQY, $\eta$) [41, 42].

$$\tau = \frac{1}{K_r + K_{nr}} \quad \text{Eq. 1}$$

$$\eta = \frac{K_r}{K_r + K_{nr}} \quad \text{Eq. 2}$$

**Figure 1(f)** shows the TRPL results of the solid-solution thin films with different Y6 loadings, where $\tau$ can be obtained by fitting the decay tail of the TRPL curves as summarized **Table S1**. We obtain a similar PL lifetime $\tau$ around 1.8 ns among those films with Y6 loadings less than 20 wt%, and it progressively decreases to around 1.3 ns with further increases in Y6 loadings, in agreement with literature value [42]. **Figure 1(g)** shows the PLQY of the solid-solution thin films with different Y6 loadings. Y6 solution has a PLQY of 20 %, while pure Y6 film has only 4 %, also consistent with the previously reported value [23]. In addition, the PLQY in the 1 wt% Y6 film is similar to that obtained in the diluted Y6 solution, further confirming that the Y6 molecules are mostly isolated in the polymer matrix. Such a large difference in PLQY between the solution and the pure film has been widely observed in organic semiconductors and is often ascribed to exciton quenching induced by molecular packing. Surprisingly, the PLQY decreases most substantially as the loading increases from 1 wt% to 20 wt%, even though the molecules in this regime remain largely isolated. Moreover, when the loading increases from 20 wt% to 100 wt%, where aggregates start to form and progressively grow, the PLQY remains nearly unchanged. Suggesting that a different mechanism is responsible for the observed PLQY reduction.

According to the TRPL and PLQY results, the $K_r$ and $K_{nr}$ values are calculated and plotted as shown in **Figure 1(h)**. Following the trend in the PLQY results as depicted in **Eq. 2**, $K_r$ decreases sharply when Y6 loading increases from 1 wt% to 20 wt%, and then plateaus at higher wt%. On the other hand, $K_{nr}$ monotonically increases as Y6 loading increases, with a higher increasing rate at the lower wt% regime. Such a steep increase in $K_{nr}$ in the low wt% regime supports the above hypothesis that the energy-transfer process is highly sensitive to the intermolecular distance between isolated molecules. This also safely excludes the contribution from Dexter charge-transfer, as it strictly requires overlapping between molecular orbitals. In addition, the value of $K_{nr}$ in the pristine Y6 film exceeds $K_r$ by more than one order of magnitude, suggesting radiative-driven photon recycling also cannot be the major energy transfer process between the Y6 molecules. It is worth noting that previous studies have attributed the increased $K_{nr}$ and the reduced PLQY in Y6 thin films, relative to solution, to molecular aggregation effects [42]. However, we find that the most pronounced changes in $K_{nr}$ and PLQY occur while the Y6 molecules remain largely isolated. Therefore, we speculated that FRET, a long-range and non-radiative energy transfer mechanism, should play a key role in mediating energy transfer between isolated molecules and aggregates in Y6 homojunction systems.

### 2.3 Role of FRET between Y6 species

#### 2.3.1 Extended FRET model for multi-component system

To investigate the contribution of FRET among different Y6 species, we first identified the donor and acceptor species and quantified their FRET efficiency. Previous studies on this process have largely been based on a semi-classical model as depicted in **Eq. 3** and **Eq. 4** [41,43].

$$E_{FRET} = \frac{1}{1+\left(\frac{r}{R_0}\right)^6} \qquad \text{Eq. 3}$$

$$R_0 = 0.0211\left(\frac{\kappa^2\,\Theta_D\,J(\lambda)}{n^4}\right)^{\frac{1}{6}} \qquad \text{Eq. 4}$$

$R_0$ represents the Förster radius, $r$ is the intermolecular separation, $\kappa^2$ denotes the orientation factor between the donor and acceptor transition dipoles, n is the refractive index of the medium, $\Theta_D$ represents the donor PLQY when acceptor is absent, and $J(\lambda)$ is the spectral overlap integral between the molar absorbance and the area-normalized emission spectrum of the donor. Direct application of this model to Y6 homojunctions faces several challenges. The conventional approach, using **Eqs. 3** and **4** to calculate FRET efficiency, considers only a single energy-transfer pathway from a specific donor species to an acceptor species. The donor-acceptor distance, $r$, is usually approximated as the intermolecular distance, assuming that the donor and acceptor molecules are homogeneously dispersed in the thin film. This approach successfully describes FRET in two-component BHJ systems [41], where donors and acceptors are typically interpenetrated on the nanoscale. However, when multiple components and energy-transfer pathways are involved, as in Y6 homojunction thin films containing various aggregate species, the overall FRET efficiency should account for both the efficiency of each individual pathway and the weight density of each component.

To resolve FRET efficiencies among distinct Y6 species, we extended the FRET model to incorporate species-dependent intermolecular separations in a multi-component system. In a system that consists of multiple FRET donor and acceptor species, the expectation value of FRET efficiency $E_{i \to j}$ between a specific donor species *i* and an acceptor species *j* can be described below:

$$E_{i \to j} = \int_a^{\infty} \frac{1}{1+\left(\frac{r}{R_{0,\, i \to j}}\right)^6} 4\pi\rho_j r^2\, e^{\left[-\frac{4}{3}\rho_j\,(r^3 - a^3)\right]}\, dr \qquad \text{Eq. 5}$$

where $\rho_j$ is the number-density of the FRET acceptor *j*, *a* is the minimum intermolecular distance between the donor and acceptor, and $R_{0,\, i \to j}$ is the corresponding Förster radius. Details of the parameterization process can be found in **SI Section 2**. $4\pi\rho_j r^2\, e^{\left[-\frac{4}{3}\rho_j\,(r^3 - a^3)\right]}$ is a 3D hard-core Poisson probability distribution function that depends on $\rho_j$, it incorporates the probability of a donor finding an acceptor with distance *r*. By integrating over *r*, it allows the determination of the expectation value of FRET efficiency $E_{i \to j}$. Hence, the overall FRET efficiency in a multi-component system can be obtained by identifying the FRET species and their corresponding number densities in the system. It is worth noting that as $\rho_j$ can be determined quantitatively by experiments in a multi-component system, it overcomes the challenge in the conventional approach as discussed above. Therefore, the framework in principle is transferable to any multi-component system in which FRET operates, provided that the individual species are spectroscopically resolvable, their donor emission and acceptor absorption overlap, and an energetic hierarchy drives downhill transfer.

### 2.3.2 Parameterizing FRET species via FCWD analysis

In parameterizing the extended FRET model, we first attempted to identify different species and estimate their number-density ratios in the solid-solution thin films to narrow the scope of the mechanistic discussion. Since the UV-Vis absorption spectrum represents the cumulative contribution of various components, FCWD analysis was applied to decompose the contributions of individual components to the overall spectral intensity. The analytical procedures follow those described in previous studies [44], and the fitting equation employed is presented below.

$$FCWD = \frac{1}{\sqrt{4\pi\lambda k_B}}\sum_{n=0}^{\infty} exp(-S)\frac{S^n}{n!}exp[-\frac{(\Delta E + n\hbar\omega + \lambda)^2}{4\lambda k_B}] \qquad \text{Eq. 6}$$

Here, $k_B$ is Boltzmann's constant, $n$ is the number of vibrational peaks, $\lambda$ is the linewidth, $S$ is the Huang-Rhys factor, $\Delta E$ is the energy difference between the vibrational energy and the 0-0 transition energy $E_{00}$, and $\hbar\omega$ represents the vibrational energy. The fitting results of several representative concentrations are shown in **Figure 2(a) to (d)**. In addition to the contributions of each component, the formation sequence of the aggregates as Y6 loading increases can also be inferred from the subsequent analysis. In the analysis, the absorption spectrum at 1 wt%, rather than that of the solution, is selected as the baseline to avoid the spectral shift effect due to different dielectric media from solution to solid, as shown in **Figure S2(a)**. The absorption profile of the non-interacting monomers (NI) derived from the 1 wt% film is shaded in yellow, while the additional peaks extracted from films with higher loadings are assigned to aggregates, as shown in **Figure 2(b)** to **2(d)**. The fitting quality across all spectra is satisfactory when assuming the presence of two types of aggregates in the Y6 films, consistent with those previously reported [44]. The contributions from aggregate I (Aggr I) and aggregate II (Aggr II) are shaded in red and blue, respectively. Due to energy-level splitting arising from intermolecular interactions [45,46], each aggregate exhibits two absorption bands at low and high energies, denoted as LE (solid line) and HE (dashed line), respectively. And the separation between the bands increases with increasing intermolecular interaction strength. Notably, the ratio between the enclosed area of each component peak and that of the whole normalized spectrum reflects not only the relative population of the corresponding species at a given loading ratio during aggregation progression, but also a more explicit physical quantity, namely the number density ratio of photons absorbed by that species upon photoexcitation (detail in **Section 2, SI**). **Figure 2(e)** summarizes how these quantities vary with Y6 loadings. Below 10 wt%, the absorption spectrum is dominated by non-interacting monomers, while aggregate II gradually increases and aggregate I remains a minor component, suggesting that the monomers initially interact with their neighboring molecules to form aggregate II as the concentration increases. Interestingly, as shown in **Figure 2(c)**, the population of aggregate I increases significantly when the Y6 loading reaches 20 wt%. This trend is more clearly illustrated in **Figure 2(e)**, which suggests that aggregate I forms abruptly at the threshold concentration of 20 wt%. With further increases in Y6 loading, the population of aggregate I continues to grow gradually, while the monomer content steadily decreases, and the aggregate II population exhibits only slight variations at higher concentrations. The proportions of different aggregates extracted by FCWD analysis in the pristine Y6 film are comparable to those obtained by molecular dynamics simulation [20,21], providing the grounds for using the FCWD results to investigate the FRET process in Y6 thin films.

The evolution of Y6 species in different loadings obtained by the FCWD analysis is consistent with the above UV-Vis absorption and PL results, confirming that the threshold behavior near 20 wt% arises from the sudden formation of a specific aggregate type. Consequently, the mechanistic discussion can be narrowed to three major components in the Y6 films, namely Y6 monomer, aggregate I, and aggregate II, to identify the potential FRET pathways between donor and acceptor species.

Regarding the potential FRET pathways, we considered fast internal exciton relaxation in each species from HE to LE bands, therefore, the monomers have the highest acceptor energy (onset at 1.52 eV), followed by the aggregate II (1.43 eV) and aggregate I (1.32 eV). Thus, three potential FRET pathways emerge: (i) from monomer to aggregate II, (ii) from monomer to aggregate I, and (iii) from aggregate II to aggregate I. When considering the monomers as the donors in the FRET calculation, the PL spectrum and PLQY of the both 1 wt% and 2 wt% film were used to determine the $R_0$ values. Whereas when considering aggregate II as donor, the PL results of the 10 wt% film were used as it has a considerable amount of aggregate II with negligible aggregate I. The number density and absorption coefficient of each species at a given loading ratio were inferred from the FCWD peak analysis. Full parameterization details and methods are given in **Section 2** of the **SI**. Although this approach may not be able to determine the exact number density of each species, it provides a sufficient semi-quantitative basis for investigating the energy transfer process, particularly for elucidating mechanistic comparisons in aggregation progression.

### 2.3.3 FRET in Y6 thin films

The parameterization described above identifies the donor–acceptor pairs associated with each FRET pathway during the aggregation progression, enabling quantification of the corresponding microscopic FRET efficiencies. Consequently, as shown in **Eq. 7**, the overall FRET efficiency, $E_{FRET}$, in a multicomponent Y6 film can be calculated as the ensemble-weighted sum of the FRET efficiencies for the individual energy-transfer pathways, where $w_{i\to j}$ denotes the ensemble weight of the corresponding FRET channel:

$$E_{FRET} = w_{M\to A2}E_{M\to A2} + w_{M\to A1}E_{M\to A1} + w_{A2\to A1}E_{A2\to A1} \qquad \text{Eq. 7}$$

As discussed above and in detail in **Section 2** of **SI**, $E_{FRET}$ is governed by both the energy-transfer distance and the spatial distribution of the FRET species within different molecular aggregates. Therefore, comparison of the aggregation-dependent FRET efficiency of each channel with the overall efficiency, $E_{FRET}$ , provides insight into how aggregation modulates microscopic FRET processes in Y6 thin films. Interestingly, as shown in **Figure 2(f)** and **Figure 1(g)**, $E_{FRFT}$ and the PLQY exhibit opposite trends with increasing Y6 loading, consistent with the non-radiative nature of the FRET process. $E_{FRFT}$ shows a sharp increase at low loadings and reaches a plateau of around 65% when the Y6 loading exceeds 10 wt%. The PLQY shows a steep initial decrease, followed by a constant value at higher Y6 loadings. The rapid increase in $E_{FRFT}$ at low loadings is due to the reduction in donor-acceptor distance $r$ towards the FRET radius $R_0$. Whereas the $E_{FRET}$ of 65% in pure Y6 film is comparable to values reported for efficient heterojunction systems [47-51]. In fact, the FRET efficiency is nearly 100% from monomers and aggregate II to aggregate I at high loadings. The maximum $E_{FRET}$ in the Y6 homojunction film is limited by the optical absorption of acceptor species, such as aggregate I, which are not involved in the transfer process. The FRET efficiency of individual channels at different loadings follows well with the amount of the corresponding FRET species, as depicted in **Figure 2(e)**. Notably, the FRET efficiency from monomer to aggregate II (M→A2) drops markedly as the Y6 loading increases from 10 wt% to 20 wt%. This is because a considerable amount of aggregate I that emerged at the 20 wt% threshold, some of the excitons generated in monomers are directly transferred to aggregate I (M→A1). It is supported by the strong spectral overlapping between the absorption of aggregate I and the PL of both monomer and aggregate II, as shown in **Figure S3**, suggesting the dominant role of aggregate I as acceptors in the FRET process in high Y6 loading films. This also accounts for the PL results in **Figure 1(d)** that the PL peaks of the 1 wt% and 2 wt% films lie at wavelengths where the neat film PL spectrum shows almost no intensity, implying that the species at low concentrations act as donors and transfer energy to the acceptors.

The above analysis provides several important insights into the energy-transfer processes in Y6 thin films. The substantial reduction in PLQY from the Y6 solution to the thin film is primarily attributed to FRET rather than to aggregation-induced exciton quenching. Moreover, the formation of various aggregates in Y6 films facilitates efficient FRET pathways. Notably, an aggregate I weight density of approximately 30-40% in Y6 films, such as in the 20 wt% Y6 film, is sufficient to achieve nearly 100% FRET efficiency from all donors. Therefore, elucidating the CT properties of these aggregates would further shed light on the origin of the strong CT characteristics observed in Y6 thin films.

### 2.3.4 Probing energy transfer and CT character between aggregates by TAS

We performed TAS measurements to investigate the energy transfer between aggregates from the dynamics of the excitation species in Y6 films with different loadings. According to previous reports [52,53], the TA spectrum of Y6 (500–1700 nm) exhibits three key features: (i) a ground-state bleaching (GSB) band at 750–850 nm corresponding to intrinsic absorption; (ii) a PIA band at around 900 nm assigned to a localized exciton (LE) with weak CT character; and (iii) an NIR signal at around 1600 nm attributed to a delocalized singlet exciton (DSE) with strong CT character.

**Figure 3(a)** shows the TA spectra of films with different Y6 loadings, probed at 0.5 ps when signal was most intense after the excitation. The curves are normalized to the minima of each GSB band. Increasing the loading from 2 wt% to 10 wt% results in only a small red shift of the GSB peak position, from around 750 to 760 nm. Further increasing to 20 wt% onwards, more pronounced shifts appear. The shift of the GSB peak is consistent with the above Y6 loading-depended UV-Vis absorption spectral characteristics.

In contrast to the pronounced redshift of the GSB, the LE exhibits only a minor red shift. The LE intensity increases with concentration up to 20 wt%, then decreases at higher concentrations, whereas the DSE is negligible below 20 wt% and increases progressively thereafter (**Figure 3(b)**). These results suggest that below 20 wt%, the film is dominated by LE species with weak CT character. Above 20 wt%, DSE species emerge and increase with concentration, and the concurrent decrease in the LE signal likely reflects the conversion of weak into strong CT species.

To further clarify the CT character and energy transfer direction, time-resolved decay dynamics of GSB, LE, and DSE were analyzed as shown in **Figure 3(c) to (e)**, respectively. Because the dynamics of GSB (**Figure 3(c)**) involves contributions from both LE and DSE, in order to understand the energy transfer process involved, the LE and DSE dynamics should be independently analyzed. LE dynamics at all concentrations show a rapid rise to a peak, followed by a steep decay (circled in **Figure 3(d)**), with higher decay rate at higher concentrations, likely due to increased LE-CT coupling [54,55]. By contrast, DSE signals are too weak to be distinguished below 20 wt%. At higher concentrations, the corresponding dynamics profiles display a delayed and slower decay after the maximum (circled in **Figure 3(e)**), consistent with an enhanced LE-CT coupling. Comparing GSB dynamics with wavelength-resolved LE and DSE dynamics reveals that below 20 wt% the GSB follows the fast LE decay feature, whereas above 20 wt% it exhibits a delayed decay characteristic of DSE. Therefore, 20 wt% is found as both the aggregation threshold and the onset of a sharp increase in CT character: monomer and aggregate II (predominant below 20%) exhibit weak CT, whereas aggregate I (which forms abruptly at 20%) displays strong CT, consistent with the reduced exciton polarizability change observed in a 20 wt% Y6 film in our prior work via electroabsorption (EA) spectroscopy [21].

We sought to elucidate the dynamics of energy transfer between species in a pure Y6 thin film (100 wt% loading) by analyzing TAS and FCWD results. Assuming excitons in higher-energy states within each species will relax to their lower-energy manifolds before the energy transfer process, only the TA dynamics at the peak positions corresponding to the lowest-energy state of each species (based on **Figure 2(d)**) were monitored. **Figure 3(f)** shows the exciton dynamics of the monomer, aggregate I, aggregate II, and DSE in a pristine Y6 thin film. The exciton densities of the monomer and aggregate II rise at nearly identical rates and reach their maxima almost simultaneously, followed by earlier decay compared to the other species. This reduction in exciton density suggests exciton migration between species, accompanied by energy transfer. Therefore, the Y6 monomer and aggregate II can be confirmed as FRET donors, as their exciton populations decay earlier according to the TAS dynamics. In addition, the instant decline in the donor profiles also indicates their weak CT character. In contrast, aggregate I exhibits a delayed rise and subsequent decay in exciton density, consistent with its role as a FRET acceptor and its stronger CT character. Similarly, DSE dynamic curve rises after the decay of the aggregate I curve. This distinctive delayed decay profile indicates that excitons, after reaching aggregate I via FRET, will subsequently relax and adopt a more delocalized character. Notably, the estimated timescale of FRET is within 1 ps as observed in **Figure 3(f)**, further confirming efficient FRET among different species in the homojunction Y6 thin film. The results from TAS agree well with the calculation results from the above FRET model. As illustrated in **Figure 3(g)**, TAS reveals that weak-CT monomers and aggregate II transfer energy to the strong-CT aggregate I, accompanied by progressive exciton delocalization.

### 2.3.5 Theoretical calculations on CT characters

So far, the major components in the Y6 film, their CT properties, and their respective roles in the FRET process have been systematically investigated. However, the geometric configurations of aggregate I and aggregate II remain unclear. To address this, we performed density functional theory (DFT) calculations to elucidate the possible packing modes of Y6 aggregates and correlate them with the aforementioned analyses. As shown in **Figure S7**, the single-crystal structure of simplified Y6 was optimized using DFT methods. Three representative dimer packing configurations, namely the S-type, C-type, and Y-type, that could potentially form aggregates were extracted and labeled according to their geometric features for clarity. These extracted dimer configurations are consistent with those reported previously [20-22].

Subsequently, additional properties of these dimer configurations, including centroid distances and interaction energies, were determined based on the optimized geometries, with detailed values provided in **Section 4** in the **SI**. Among the identified configurations, the S dimer exhibits the largest centroid distance (1.9 nm), suggesting that it is preferentially formed at the early stage of aggregation. In contrast, the C and Y dimers have much smaller centroid distances (0.8 nm); therefore, they are expected to form at later stages of aggregation in films with high Y6 concentration. Since the interaction energy of the Y dimer is significantly larger than that of the C dimer, the C dimer is expected to have the lowest population, which can be attributed to its lower thermodynamic stability and the greater requirement for close intermolecular contact. Consequently, aggregate II likely corresponds to the S dimer, which forms earlier, while aggregate I is associated with the Y dimer, occurring at a later aggregation stage.

To further substantiate these assignments and provide deeper insight into the CT characteristics of these species, the excited-state properties of the three major species in the Y6 film were investigated using time-dependent density functional theory (TDDFT) calculations on their first five excited states, as detailed in the **SI**. Hole-electron wavefunction analysis was performed quantitatively based on the TDDFT results, as shown in **Table 1**. The *Sr* reflects the overlap between the electron and hole centroids; a smaller *Sr* indicates greater spatial separation and stronger CT character. The *D* index measures the centroid distance between electron and hole. A larger *D* implies that the electron and hole centroids are more widely separated in space and therefore enhanced CT properties [56]. The *t* index quantifies the degree of electron-hole spatial separation, with more positive values corresponding to stronger CT characteristics. Therefore, consistent with previous reports [20-24,54,55], the Y dimer exhibits the strongest CT character, indicating a greater propensity to form delocalized excitons. In contrast, the S-type dimer and the Y6 monomer display weaker CT character, suggesting a tendency to form more localized excitons. Based on these results, we conclude that the S dimer corresponds to aggregate II, which forms earlier during the aggregation process and exhibits weak CT character, whereas the Y-type dimer corresponds to aggregate I, which forms at later stages and is characterized by strong CT character.

## 3. Physical insight into aggregate-mediated energy funneling in NFA thin film

The above calculation of the FRET efficiency in the Y6 thin film assumes only single-step energy transfer between species. However, as shown in **Figure S3(c)**, the extinction coefficient of the Y-type aggregate strongly overlaps with the PL spectra of both donor species. Therefore, in addition to direct exciton transfer from monomers to Y-type aggregates, excitons initially transferred from monomers to S-type aggregates may subsequently transfer to Y-type aggregates and form DSE after relaxation. Specifically, we quantify the collection efficiency of excitons that ultimately reach the Y-type aggregates, $E_{collection}$, as defined in **Eq. S12**. It represents the probability that an exciton is either initially generated directly on Y-type aggregates or arrives at them via FRET. The dependence of $E_{collection}$ and $E_{FRET}$ on aggregation is shown in **Figure 4(a)**. Below 20 wt%, the thin film is dominated by monomer and S-type aggregates, with FRET proceeding mainly through the monomer to S-type aggregate channel. As a result, only a few excitons are initially generated or transferred via FRET on Y-type aggregates. Consequently, even $E_{\mathrm{REFT}}$ can reach above 50%, $E_{collection}$ remains below 20%. Excitons are therefore predominantly distributed among the weak CT species (monomer and S-type aggregates), and the film exhibits only weak CT character. Above 20 wt%, $E_{collection}$ jumps sharply to larger than 80%. This sharp rise stems not only from the increase of $E_{\mathrm{FRET}}$ due to the abrupt increase of Y-type aggregates population, but also from the increase of excitons directly generated in Y-type aggregates. Taken together, the FRET-mediated and directly excited contributions explain the strong CT character in the pure Y6 film, even though there are still considerable numbers of weak CT species, in agreement with the TAS and EA results described above [21].

**Figures 4(b) and 4(c)** summarize the above physical picture for neat Y6 films. Upon photon absorption, the Y6 film initially generates localized excitons on Y6 monomers or S-type aggregates, both of which possess weak CT characteristics and act as FRET donors. With intrinsic downhill transfer probability near unity, these donors subsequently transfer energy to acceptors, namely the Y-type aggregates exhibiting stronger CT features, via ultrafast (<1 ps) and efficient FRET process, accompanied by exciton diffusion. Eventually, the excitons develop more pronounced CT character, forming DSEs that tend to dissociate at the acceptor sites.

The key physical insight revealed here is that the molecular structures of Y6 uniquely enable the formation of multiple species, including monomers and various aggregate configurations in thin films. These species possess distinct photophysical characteristics that drive energy transfer and exciton diffusion from weak CT species to stronger CT counterparts via an efficient and ultrafast FRET process.

The physical picture established here for Y6 is not specific to this molecule. It relies on three general conditions that are widely satisfied in NFA thin films: (i) the coexistence of multiple packing motifs with distinct CT characters, (ii) spectral overlap between the emission of weak CT species and the absorption of strong CT species, and (iii) an energetic hierarchy that drives downhill energy transfer. These conditions arise naturally from the A-D-A′–D-A architecture of Y-series acceptors. Therefore, the multi-component FRET framework developed here provides a general tool to quantify exciton funneling in any NFA film, and the resulting weak CT to strong CT energy cascade can be regarded as a general design principle for organic photovoltaics without nanoscale BHJ. This fundamentally explains why relatively high EQE can still be achieved in such devices, as well as the origin of strong overall CT character of the Y6 thin film [57,58], even when a considerable population of weakly CT isolated molecules exists [59].

## 4. Conclusion

In conclusion, through comprehensive experimental characterization, modeling, and theoretical calculations, we have demonstrated for the first time that FRET serves as a vital long-range energy-transfer mechanism between molecular aggregates in Y6 homojunctions. By accounting for multiple parallel energy-transfer channels among different Y6 species, we observe a high FRET efficiency of 65% on a sub-picosecond timescale (<1 ps) in Y6 homojunction thin films, comparable to that reported for conventional donor–acceptor heterosystems. This efficient FRET is attributed to the coexistence of multiple aggregate species in Y6 homojunctions. Aggregates with complementary CT character act as FRET donors and acceptors, enabling efficient energy transfer and facilitating both DSE formation and exciton dissociation. Our findings establish a general mechanistic principle that weak CT species act as donors that directionally funnel energy to strong CT acceptors via ultrafast FRET. These findings provide a unified mechanistic explanation for the recently reported outstanding performance of Y6-based single-component and low-donor-ratio heterojunction devices. Ultimately, this work offers new insights into leveraging molecular design and molecular packing to advance the development of next-generation organic electronic devices.

## 5. Acknowledgement

We gratefully acknowledge financial support from the National Natural Science Foundation of China (NSFC, Grant No. 62474151). We also acknowledge Prof. Andrey L. Rogach (Department of Materials Science and Engineering, City University of Hong Kong) for access to the Edinburgh FLS1000 photoluminescence spectrometer.

## 6. References

[1]. Tang, C. W. Two-layer organic photovoltaic cell. *Applied physics letters* **48**, 183-185 (1986).

[2]. Yu, G., Gao, J., Hummelen, J. C., Wudl, F. & Heeger, A. J. Polymer photovoltaic cells: enhanced efficiencies via a network of internal donor-acceptor heterojunctions. *Science* **270**, 1789-1791 (1995).

[3]. Liu, J. et al. Fast charge separation in a non-fullerene organic solar cell with a small driving force. *Nature Energy* **1**, 16089 (2016).

[4]. Tang, C. W. & VanSlyke, S. A. Organic electroluminescent diodes. *Applied physics letters* **51**, 913-915 (1987).

[5]. Uoyama, H., Goushi, K., Shizu, K., Nomura, H. & Adachi, C. Highly efficient organic light-emitting diodes from delayed fluorescence. *Nature* **492**, 234-238 (2012).

[6]. Yu, G., Pakbaz, K. & Heeger, A. Semiconducting polymer diodes: Large size, low cost photodetectors with excellent visible-ultraviolet sensitivity. *Applied Physics Letters* **64**, 3422-3424 (1994).

[7]. Xing, S. et al. Photomultiplication-type organic photodetectors for near-infrared sensing with high and bias-independent specific detectivity. *Advanced Science* **9**, 2105113 (2022).

[8]. Hou, J., Inganäs, O., Friend, R. H. & Gao, F. Organic solar cells based on non-fullerene acceptors. *Nature materials* **17**, 119-128 (2018).

[9]. Lin, Y. *et al.* An electron acceptor challenging fullerenes for efficient polymer solar cells. *Advanced materials* **27**, 1170-1174 (2015).

[10]. Cui, C. Recent progress in fused-ring based nonfullerene acceptors for polymer solar cells. *Frontiers in Chemistry* **6**, 404 (2018).

[11]. Yuan, J. *et al.* Single-junction organic solar cell with over 15% efficiency using fused-ring acceptor with electron-deficient core. *Joule* **3**, 1140-1151 (2019).

[12]. Wang, J., Zhao, W., Bai, Y., Long, G. & Kan, B. Mitigating Nonradiative

Recombination Loss Using Alkoxy‐Substituted Third‐Component Acceptor for 20.1%‐Efficiency Organic Solar Cell. *Advanced Functional Materials*, e14789 (2025).

[13]. Cao, X. *et al.* O, S, and N Bridged Atoms Screening on 2D Conjugated Central Units of High-Performance Acceptors. *Advanced Materials*, 2503131 (2025).

[14]. Zhu, Y. *et al.* Regulating H-/J-Aggregation and Suppressing Exciton-Vibration Coupling of Fused Ring Electron Acceptors Enable 20.51% Efficiency of Organic Solar Cells. *Advanced Materials*, e17029 (2025).

[15]. Han, G. & Yi, Y. Molecular insight into efficient charge generation in low-driving-force nonfullerene organic solar cells. *Accounts of Chemical Research* **55**, 869-877 (2022).

[16]. Bakulin, A. A. *et al.* The role of driving energy and delocalized states for charge separation in organic semiconductors. *Science* **335**, 1340-1344 (2012).

[17]. Fan, H. *et al.* Side-Chain Controlled Exciton Binding Energy in Y-Acceptors Toward Homo-Junction Organic Solar Cells With 5.07% Efficiency. *Small*, e13943 (2025).

[18]. Dolan, A. *et al.* Enhanced photocatalytic and photovoltaic performance arising from unconventionally low donor–Y6 ratios. *Advanced Materials* **36**, 2309672 (2024).

[19]. Yao, N. *et al.* Efficient charge transport enables high efficiency in dilute donor organic solar cells. *The Journal of Physical Chemistry Letters* **12**, 5039-5044 (2021)

[20]. Kupgan, G., Chen, X.-K. & Brédas, J.-L. Molecular packing of non-fullerene acceptors for organic solar cells: Distinctive local morphology in Y6 vs. ITIC derivatives. *Materials Today Advances* **11**, 100154 (2021).

[21]. Mahadevan, S. *et al.* Assessing intra-and inter-molecular charge transfer excitations in non-fullerene acceptors using electroabsorption spectroscopy. *Nature Communications* **15**, 2393 (2024).

[22]. Han, G., Hu, T. & Yi, Y. Reducing the singlet− triplet energy gap by end-group

π− π stacking toward high-efficiency organic photovoltaics. *Advanced Materials* **32**, 2000975 (2020).

[23]. Zhang, G. *et al.* Delocalization of exciton and electron wavefunction in non-fullerene acceptor molecules enables efficient organic solar cells. *Nature communications* **11**, 3943 (2020).

[24]. Giannini, S. *et al.* On the role of charge transfer excitations in non-fullerene acceptors for organic photovoltaics. *Materials Today* **80**, 308-326 (2024)

[25]. Bowman, A. R., Anaya, M., Greenham, N. C. & Stranks, S. D. Quantifying photon recycling in solar cells and light-emitting diodes: absorption and emission are always key. *Physical review letters* **125**, 067401 (2020).

[26]. Monguzzi, A., Tubino, R. & Meinardi, F. Upconversion-induced delayed fluorescence in multicomponent organic systems: Role of Dexter energy transfer. *Physical Review B—Condensed Matter and Materials Physics* **77**, 155122 (2008).

[27]. Cravcenco, A., Ye, C., Gräfenstein, J. r. & Börjesson, K. Interplay between Förster and Dexter Energy Transfer Rates in Isomeric Donor–Bridge–Acceptor Systems. *The Journal of Physical Chemistry A* **124**, 7219-7227 (2020).

[28]. Cnops, K. *et al.* 8.4% efficient fullerene-free organic solar cells exploiting long-range exciton energy transfer. *Nature communications* **5**, 3406 (2014).

[29]. Menke, S. M. & Holmes, R. J. Exciton diffusion in organic photovoltaic cells. *Energy & Environmental Science* **7**, 499-512 (2014).

[30]. Mohapatra, A. A. *et al.* Role of Morphology and Förster Resonance Energy Transfer in Ternary Blend Organic Solar Cells. *Acs Applied Energy Materials* **3**, 12025-12036 (2020).

[31]. Bi, P. *et al.* Dual Förster resonance energy transfer effects in non-fullerene ternary organic solar cells with the third component embedded in the donor and acceptor. *Journal of materials chemistry A* **5**, 12120-12130 (2017).

[32]. Firdaus, Y. *et al.* Long-range exciton diffusion in molecular non-fullerene acceptors. *Nature communications* **11**, 5220 (2020).

[33]. Hodgkiss, J. M. *et al.* Exciton-Charge Annihilation in Organic Semiconductor Films. *Advanced Functional Materials* **22**, 1567-1577 (2012).

[34]. Zhang, K.-N. *et al.* Exploring the mechanisms of exciton diffusion improvement in ternary polymer solar cells: From ultrafast to ultraslow temporal scale. *Nano Energy* **79**, 105513 (2021).

[35]. Jang, Y. J. *et al.* Enhancing the organic solar cell efficiency by combining plasmonic and Förster Resonance Energy Transfer (FRET) effects. *Journal of Power Sources* **438**, 227031 (2019).

[36]. Chandrabose, S. *et al.* High exciton diffusion coefficients in fused ring electron acceptor films. *Journal of the American Chemical Society* **141**, 6922-6929 (2019).

[37]. Park, S. Y. *et al.* Photophysical pathways in efficient bilayer organic solar cells: The importance of interlayer energy transfer. *Nano Energy* **84**, 105924 (2021).

[38]. Menke, S. M., Luhman, W. A. & Holmes, R. J. Tailored exciton diffusion in organic photovoltaic cells for enhanced power conversion efficiency. *Nature materials* **12**, 152-157 (2013).

[39]. Sayner, T., Ruseckas, A., Harwell, J. R. & Samuel, I. D. Exciton Diffusion to Low Energy Sites of the Acceptor Drives Charge Photogeneration in D18: Y6 Solar Cells. *The Journal of Physical Chemistry C* **128**, 19319-19328 (2024).

[40]. Scholes, G. D. Long-range resonance energy transfer in molecular systems. *Annual review of physical chemistry* **54**, 57-87 (2003).

[41]. Lakowicz, J. R. *Principles of fluorescence spectroscopy.* (Springer, 2006).

[42]. Wang, Z. *et al.* The role of interfacial donor–acceptor percolation in efficient and stable all-polymer solar cells. *Nature Communications* **15**, 1212 (2024).

[43]. Benatto, L. *et al.* Fret–calc: A free software and web server for förster resonance energy transfer calculation. *Computer Physics Communications* **287**, 108715 (2023).

[44]. Kroh, D. *et al.* Identifying the signatures of intermolecular interactions in blends of PM6 with Y6 and N4 using absorption spectroscopy. *Advanced Functional*

*Materials* **32**, 2205711 (2022).

[45]. Kasha, M. Energy transfer mechanisms and the molecular exciton model for molecular aggregates. *Radiation research* **20**, 55-70 (1963).

[46]. Kasha, M., Rawls, H. R. & El-Bayoumi, M. A. The exciton model in molecular spectroscopy. *Pure and applied Chemistry* **11**, 371-392 (1965).

[47]. Jin, J. *et al.* Recent developments of polymer solar cells with photovoltaic performance over 17%. *Advanced Functional Materials* **33**, 2213324 (2023).

[48]. Wang, T., Wang, X., Yang, R. & Li, C. Recent advances in ternary organic solar cells based on förster resonance energy transfer. *Solar RRL* **5**, 2100496 (2021).

[49]. Chen, Z. *et al.* Ultrafast energy transfer from polymer donors facilitating spectral uniform photocurrent generation and low energy loss in high-efficiency nonfullerene organic solar cells. *Energy & Environmental Science* **16**, 3373-3380 (2023).

[50]. Zhang, K. *et al.* Natural Photosensitizer for Regulating Organic Solar Cells: Coupling Mechanisms of Intermolecular Interactions and Energy-Charge Transfer. *The Journal of Physical Chemistry Letters* (2026).

[51]. Zhou, M. *et al.* Improved Exciton Diffusion through Modulating Förster Resonance Energy Transfer for Efficient Organic Solar Cells. *Solar RRL* **8**, 2400136 (2024).

[52]. Natsuda, S.-i. *et al.* Singlet and triplet excited-state dynamics of a nonfullerene electron acceptor Y6. *The Journal of Physical Chemistry C* **125**, 20806-20813 (2021).

[53]. Tamai, Y., Murata, Y., Natsuda, S. i. & Sakamoto, Y. How to interpret transient absorption data?: an overview of case studies for application to organic solar cells. *Advanced Energy Materials* **14**, 2301890 (2024).

[54]. Qian, D. *et al.* Correlating the hybridization of local-exciton and charge-transfer states with charge generation in organic solar cells. *Advanced Energy Materials* **13**, 2301026 (2023).

[55]. Coropceanu, V., Chen, X.-K., Wang, T., Zheng, Z. & Brédas, J.-L.

Charge-transfer electronic states in organic solar cells. *Nature Reviews Materials* **4**, 689-707 (2019).

[56]. Liu, Z., Lu, T. & Chen, Q. An sp-hybridized all-carboatomic ring, cyclo [18] carbon: Electronic structure, electronic spectrum, and optical nonlinearity. *Carbon* **165**, 461-467 (2020).

[57]. Jiang, K. *et al.* Pseudo-bilayer architecture enables high-performance organic solar cells with enhanced exciton diffusion length. *Nature communications* **12**, 468 (2021).

[58]. Liu, S. *et al.* High-Performance Pseudo-Bilayer Organic Solar Cells Enabled by Sequential Deposition of D18/Y6 Chloroform Solution. *ACS Applied Energy Materials* **6**, 5047-5057 (2023).

[59]. Zuo, L. *et al.* Dilution effect for highly efficient multiple-component organic solar cells. *Nature nanotechnology* **17**, 53-60 (2022).

# 7. Figures & table

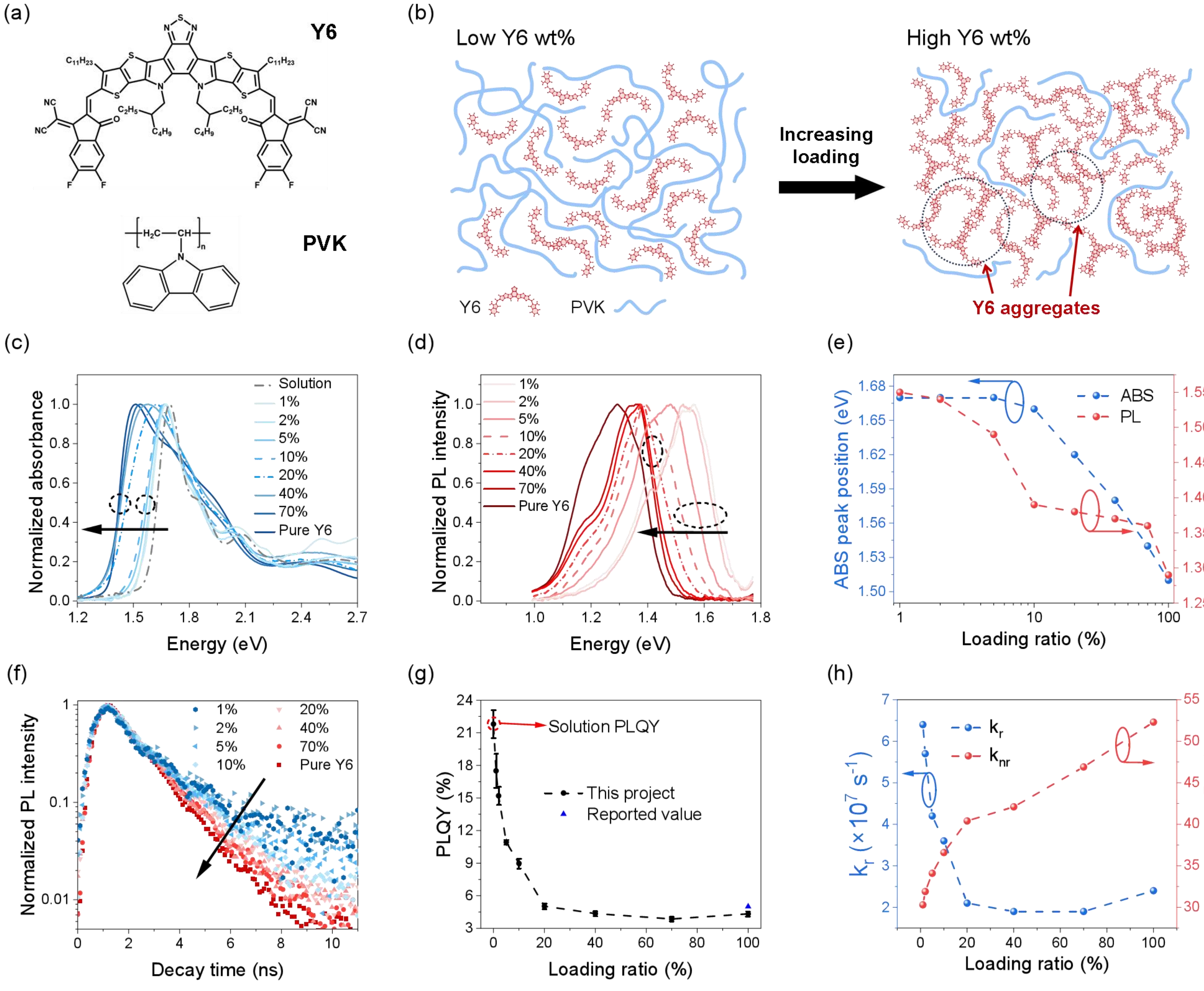

**Figure 1**. **Identification of potential energy transfer mechanisms in Y6 thin film via tuning the aggregation.** (a) Chemical structure of Y6 and PVK; (b) As the loading ratio increases, Y6 molecules dispersed in the PVK matrix form aggregate. The PVK molecules are represented by red curved lines; Corresponding (c) UV-vis absorption, (d) PL spectra with (e) Peak positions, (f) TRPL spectra, (g) PLQY and (h)

Calculated $K_r$ and $K_{nr}$ of Y6:PVK films with different loading ratios. The evolution trend with increasing loading ratio is indicated by black arrows and the features of absorption and PL edges in low and high concentration regions are highlighted by dashed circles.

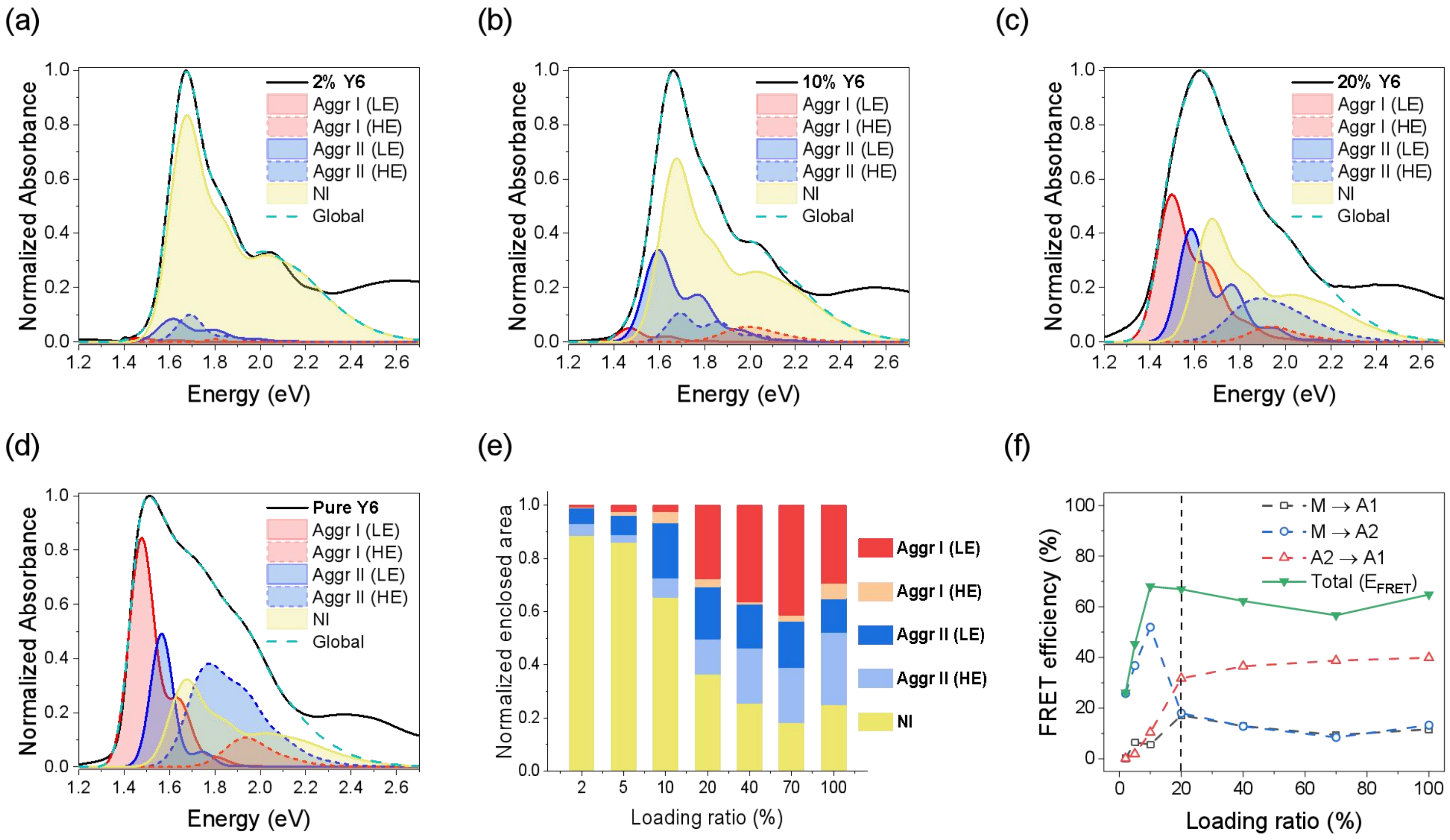

**Figure 2. FCWD analysis to investigate major species in Y6 thin film for mechanistic discussion.** UV-Vis absorption spectra decomposition of (a) 2 wt% ; (b) 10 wt% ; (c) 20 wt% and (d) Pure Y6 thin films by FCWD method. Three major species aggregate I/II (Aggr I/II) and non-interacting monomer (NI) are denoted by red, blue and yellow in the fitted peaks, respectively. The absorption edge with lower energy (LE) is denoted by the solid line while that with higher energy (HE) is denoted by dashed line for aggregates. The relative contribution of each species to the general absorbance is represented by the enclosed area of each peak; (e) The summary of the variation of species relative populations (by enclosed area) and (f) Total and each pathway effective FRET efficiency with different loading ratios. M, A1 and A2

represents monomer, aggregate I and II, respectively and the arrow indicates the energy transfer direction

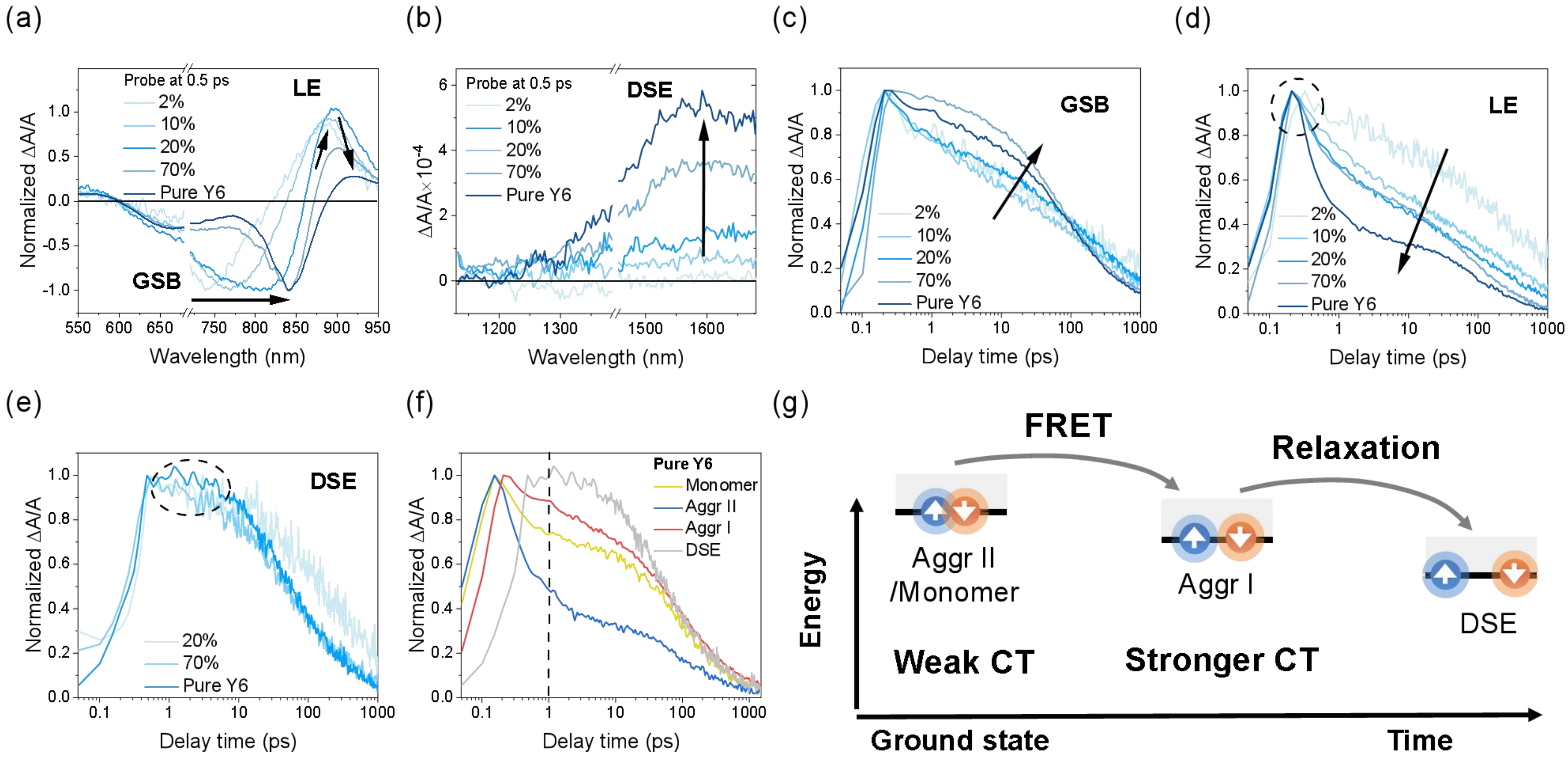

**Figure 3**. **TAS and decay dynamics to probe CT properties of species and identify FRET donor/acceptor with directional energy transfer pathway in Y6 thin film.** (a) TA spectra normalized to GSB at 0.5 ps and (b) NIR TA spectra of Y6:PVK thin films with different loading ratios. The evolution of GSB, LE and DSE features with increasing concentration is indicated by arrows; TA decay dynamics of (c) GSB (around 750-850 nm, minima of each curve); (d) LE (around 900 nm); (e) DSE (around 1600 nm) and (f) Major species in Y6 film probed at corresponding feature wavelengths based on FCWD analysis and TAS (monomer: 753 nm, aggregate I: 837 nm, aggregate II: 796 nm and DSE: 1595 nm). The variation of features with increasing loading ratio is shown by arrows, and the difference in decay dynamics between LE and DSE is highlighted by dashed circles; (g) Schematic of FRET directional pathway from weak CT donor (monomer/aggregate II) to stronger CT acceptor (aggregate I), followed by formation of DSE after relaxation.

**Table 1**. Wavefunction analysis result

| Species in FRET model | Corresponding species in Y6 thin film | *Sr* (a.u.) | *D* (Å) | *t* (Å) |
|---|---|---|---|---|
| NI (M) | Monomer | 0.693 | 1.069 | -3.602 |
| Aggr II (A2) | Dimer S | 0.681 | 0.009 | -6.374 |
| / | Dimer C | 0.678 | 0.899 | -4.908 |
| Aggr I (A1) | Dimer Y | 0.671 | 2.799 | -2.961 |

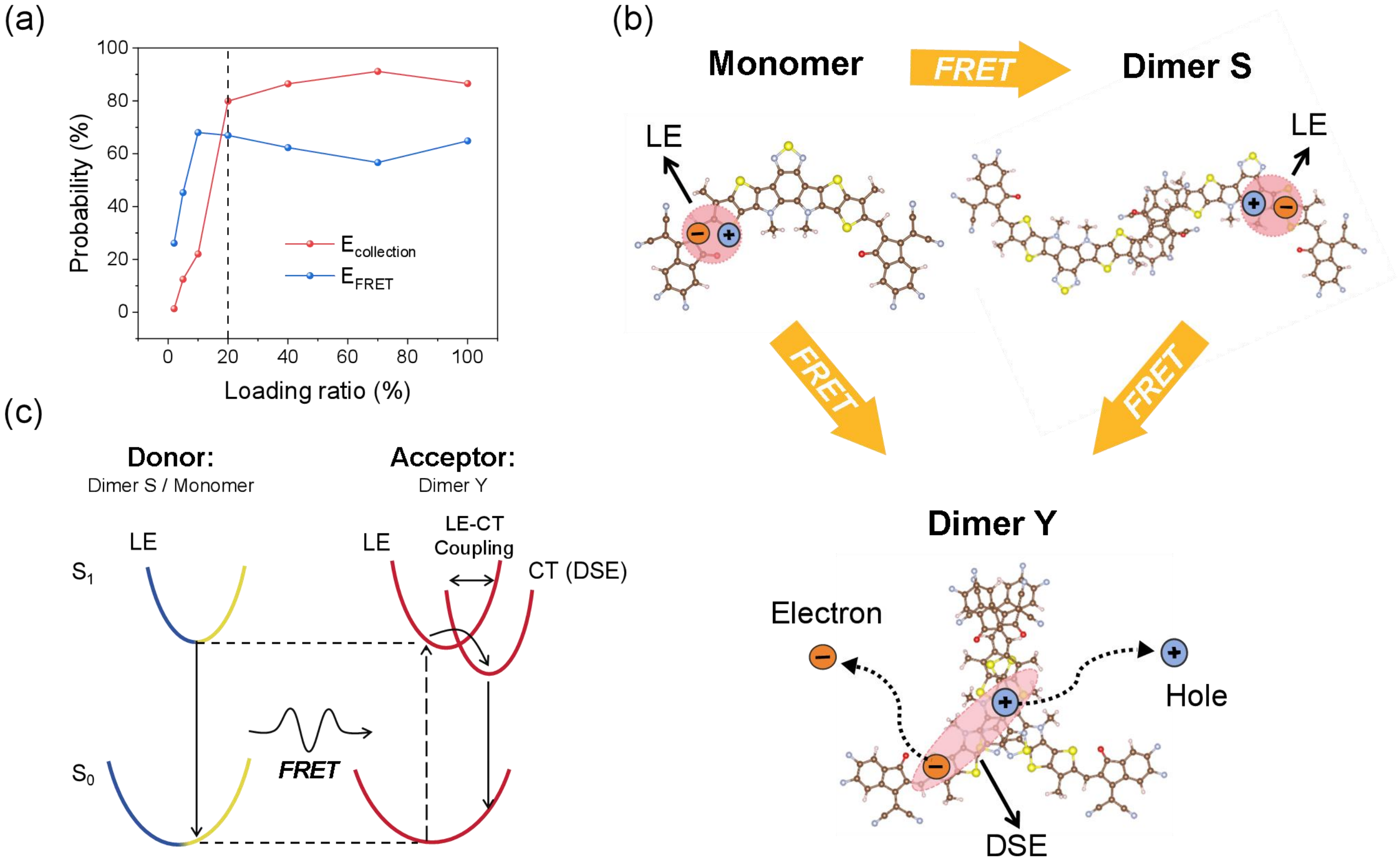

**Figure 4**. **Correlation of FRET and CT properties with schematic of FRET process in Y6 homojunction.** (a) Comparison of aggregation dependent FRET efficiency and exciton collection efficiency on Y-type aggregate;(b) From geometry: LE on the FRET donor (monomer or S-type aggregate), which finally diffuses to the acceptor species (Y-type aggregate) via FRET. The subsequent relaxation forms DSE with a tendency to dissociate; (c) From energy and dynamics: The exciton from donor species with weak CT property diffuses to stronger CT acceptor species within 1 ps.

# Supplementary Information

# Ultrafast Inter-Aggregate Energy Transfer Drives Efficient Charge Generation in Organic Homojunctions

*Shanchao OUYANG[1,2], Hang Yuen HO[1,2], Sudhi MAHADEVAN[1,2], Ziyao YUE[1,2], Haochen LIU[1], Zhiqiang GUAN[2,3], Chun-Sing LEE[2,3], Shu Kong SO[4] and Sai-Wing TSANG[1,2,5*]*

1 Department of Materials Science and Engineering, City University of Hong Kong, Hong Kong SAR, PR China

2 Centre of Super-Diamond and Advanced Films, City University of Hong Kong, Hong Kong SAR, PR China.

3 Department of Chemistry, City University of Hong Kong, Hong Kong SAR, PR China

4 Department of Physics and Institute of Advanced Materials, Hong Kong Baptist University, Kowloon Tong, Hong Kong SAR, PR China

5 The City University of Hong Kong Shenzhen Research Institute, Shenzhen, PR China

Email:saitsang@cityu.edu.hk

# List of Content

## Section 1. TRPL & PLQY result

To quantify the radiative and non-radiative recombination rates, we first determined the PL lifetime for each sample by performing a bi-exponential fit to the decay region of the TRPL profiles. The corresponding fitting curves are presented in **Figure S1**. Subsequently, the PLQY of each sample was measured. Based on the obtained PL lifetime and PLQY values, the radiative and non-radiative decay rates were calculated. The variation of these parameters (PL lifetime $\tau$ , PLQY $\eta$, $K_r$ and $K_{nr}$) was summarized in **Table S1** as a function of the loading ratio, which serves as the source data for **Figure 1(h)** in the main text.

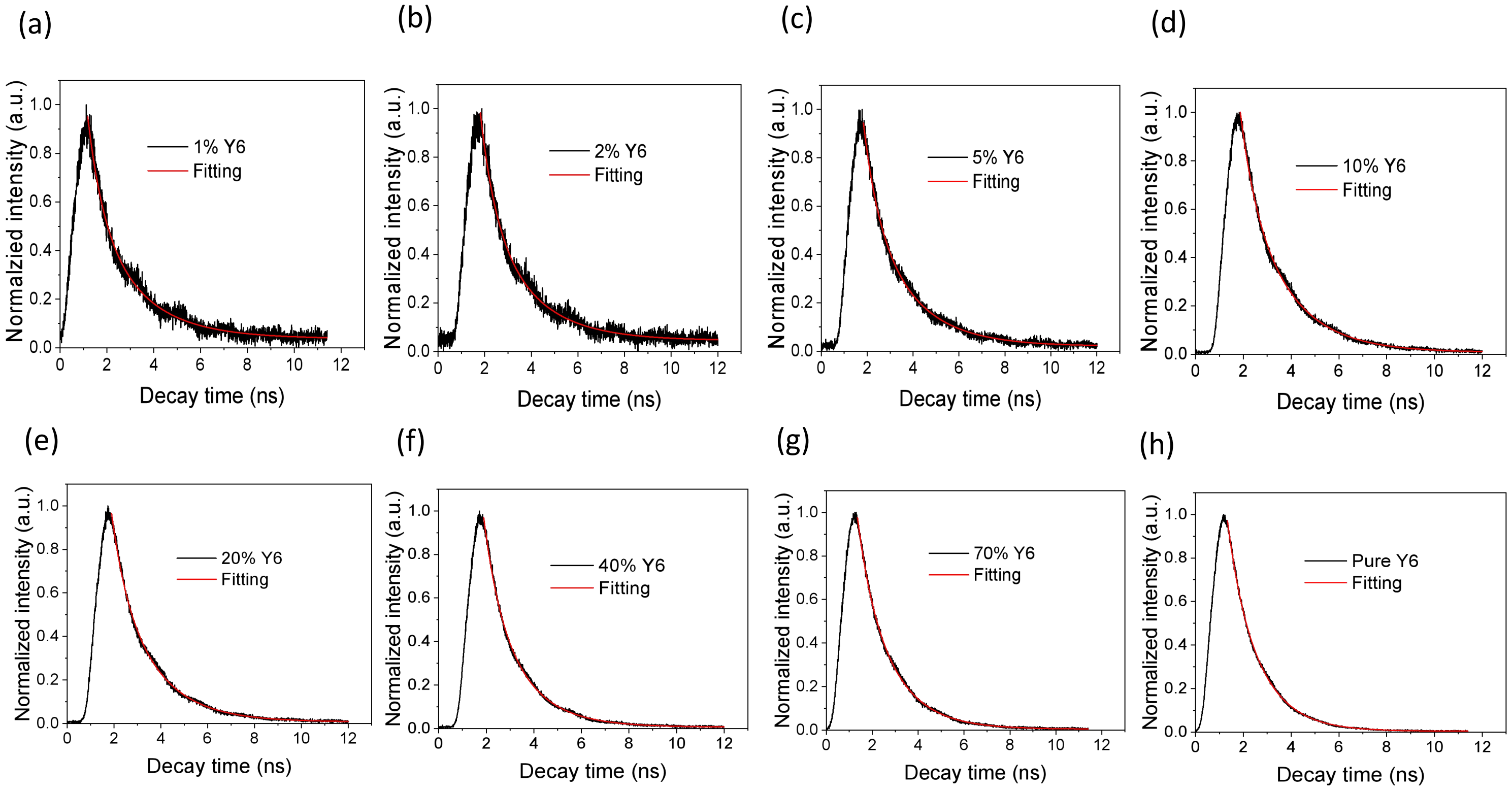

**Figure S1.** Fitting of TRPL for (a) 1 wt%; (b) 2 wt%; (c) 5 wt%; (d) 10 wt%; (e) 20 wt%; (f) 40 wt%; (g) 70 wt% and (h) pure Y6 in PVK thin films

**Table S1.** Summary for $K_r$ and $K_{nr}$ calculation

| **Loading ratio (%)** | $\tau$ **(ns)** | $\eta$ **(%)** | $K_r$ **(×10$^7$ s$^{-1}$)** | $K_{nr}$ **(×10$^7$ s$^{-1}$)** |
|---|---|---|---|---|
| 1 | 1.84 | 17.5 | 9.52 | 44.9 |
| 2 | 1.78 | 15.2 | 8.54 | 47.6 |
| 5 | 1.77 | 10.9 | 6.16 | 50.3 |
| 10 | 1.75 | 8.98 | 5.13 | 51.9 |
| 20 | 1.64 | 5.04 | 3.05 | 58.0 |
| 40 | 1.44 | 4.36 | 3.05 | 66.3 |
| 70 | 1.43 | 3.86 | 2.74 | 67.5 |
| 100 | 1.28 | 4.33 | 3.36 | 74.7 |

## Section 2. FRET model parameterization and results

The theory describing FRET involves a detailed framework derived from both classical electrodynamics and quantum mechanical principles. For practical application, a semi-classic model is often employed; thus, only the key governing equations are presented in the main text [1, Ref 41 in main text]. Fundamentally, FRET arises from dipole-dipole interactions between an excited donor and a ground state acceptor. Its efficiency, denoted $E_{FRET}$, depends critically on the intermolecular separation $r$ and is expressed in **Eq. 3** and **Eq. 4**.

In detail, $R_0$ represents the Förster radius, defined as the donor-acceptor separation at which the efficiency of FRET is 50%. In its expression, for a randomly oriented system such as a thin film, $\kappa^2$ is conventionally assumed to be 2/3 and the value of donor $\Theta_D$ is determined experimentally via PLQY measurements. To determine $R_0$ and the efficiency of any energy transfer channel through **Eq. 5**, the spectral overlap integral $J(\lambda)$, the dielectric constant $n$, and the parameter $a$ under the corresponding conditions are required.

To estimate the minimum intermolecular distance $a$, we assume a uniform distribution of molecules within the thin film. Each molecule occupies a cubic volume of 1 $cm^3$. The molar mass M of the Y6 molecule is approximately 1451.93 g/mol (as provided by the chemical supplier). Assuming the densities of Y6 and PVK are both 1.2 g/$cm^3$, we define $N$ as the molar number, $V_M$ as the molar volume, and $N_A$ is Avogadro's number. In the pure Y6 system:

$$N = \frac{\sigma N_A}{M}, V_M = \frac{1}{N} \quad \text{Eq. S1}$$

Given that the molar volume $V_M$ corresponds to the volume occupied by one mole of molecules in this cubic lattice model, the following relationship holds:

$$a = (V_M)^{\frac{1}{3}} = 1.26\ (nm) \quad \text{Eq. S2}$$

The refractive index $n$ in **Eq. 4** is approximated as the square root of the dielectric constant. The dielectric constants for Y6 and PVK are taken as 4 and 3, respectively [2,3]. Based on the cubic molecular packing model established above, the effective volume occupied by each Y6 molecule is comparable to that of a sphere with radius $r$. Consequently, the effective dielectric constant for a given loading ratio can be estimated using the Maxwell-Garnett effective medium approximation [4,5].

For the spectral overlap integral $J(\lambda)$, the area normalized PL spectrum of donor and molar absorption coefficient of the acceptor are required. The coefficient depends on the film density, molar mass and thickness. The PL spectrum can be measured directly, whereas the absorption coefficient is derived from the UV-Vis absorption spectrum via the Beer-Lambert law (**Eq. S3**). In this expression, $A(\lambda)$ is the absorbance at wavelength $\lambda$ (nm), $\varepsilon(\lambda)$ is the corresponding absorption coefficient, $c$ is the molar concentration of the species of interest (loading ratio of each sample), and $l$ is the optical path length, approximated as the film thickness, which is 100 nm for all samples.

$$A(\lambda) = \varepsilon(\lambda) \cdot c \cdot l \qquad \text{Eq. S3}$$

As mentioned in the main text, before calculating the efficiency of a given energy transfer channel, the donor and acceptor species and the number density of the acceptors must be determined through FCWD analysis. From the density, molar mass, and loading ratio of Y6, the number density of Y6 molecules in the sample can be obtained. The number density of each species can then be derived from their respective fractions in the FCWD analysis. The extinction coefficient spectrum of each species can also be obtained via **Eq. S3** and the relative intensity ratio between the component peaks and the global peak in the FCWD analysis.

Notably, in this procedure, we estimated the acceptor number density by multiplying the total Y6 number density at a given loading ratio by the ratio of the FCWD acceptor peak area to the total spectral area. However, this area ratio is not strictly equivalent to the true fractional number density of the acceptor species. According to **Eq. S3**, the ratio $\phi_i$ of the area enclosed by species i to the total spectral area can be written in the form of **Eq. S4**.

$$\phi_i = \frac{A_i(\lambda)\, d\lambda}{\sum A(\lambda)\, d\lambda} = \frac{\int_0^\infty \varepsilon_i(\lambda)\cdot c_i \cdot l \cdot \lambda d\lambda}{\sum \int_0^\infty \varepsilon(\lambda)\cdot c \cdot l \cdot \lambda d\lambda} \qquad \text{Eq. S4}$$

Assuming a uniform film thickness $l$ across all samples, the final expression for $\phi_i$ is given by **Eq. S5**. $\phi_i$ depends on both the relative number density and the extinction coefficient of species $i$. Physically, it represents the ratio of the photon number density absorbed by species $i$ to the total absorbed photon number density, that is, the relative initial exciton number density generated on that species.

$$\phi_i = \frac{\int_0^\infty \varepsilon_i(\lambda)\cdot c_i\cdot\lambda d\lambda}{\sum\int_0^\infty \varepsilon(\lambda)\cdot c\cdot\lambda d\lambda} \qquad \text{Eq. S5}$$

Because the extinction coefficient is proportional to the oscillator strength, the area ratio $\phi_i$ approximates the product of the number density fraction of species $i$ and the corresponding oscillator strength ratio. In the Y6 system, the oscillator strengths of different species are within the same order of magnitude [Ref 21-24 in main text], so the discrepancy between the actual number density fraction and the exciton number density fraction is expected to be small. Thus, using this area to estimate the number densities of individual species should be qualitatively correct in terms of their relative magnitudes. The ratio of species derived from this approach also consistent with that determined by molecular dynamic simulation [Ref 21 in main text]. Furthermore, because a central goal of this study is to characterize relative trends during aggregation progression, and our main conclusions are drawn from these trends, we therefore consider the error introduced by this approximation acceptable.

**Figure. S2** presents the FCWD fitting results for the 1 wt%, 5 wt%, 40 wt% and 70 wt% Y6:PVK thin film samples. Together with the loading ratios reported in the main text, the number density of any species, the exciton distribution, and the extinction coefficient at all concentrations can be parameterized via spectroscopic analysis. To minimize artifacts from solution-to-solid spectral shifts, the absorption spectrum of the 1 wt% thin film (rather than the solution spectrum) was used as the fixed baseline for the fitting procedure.

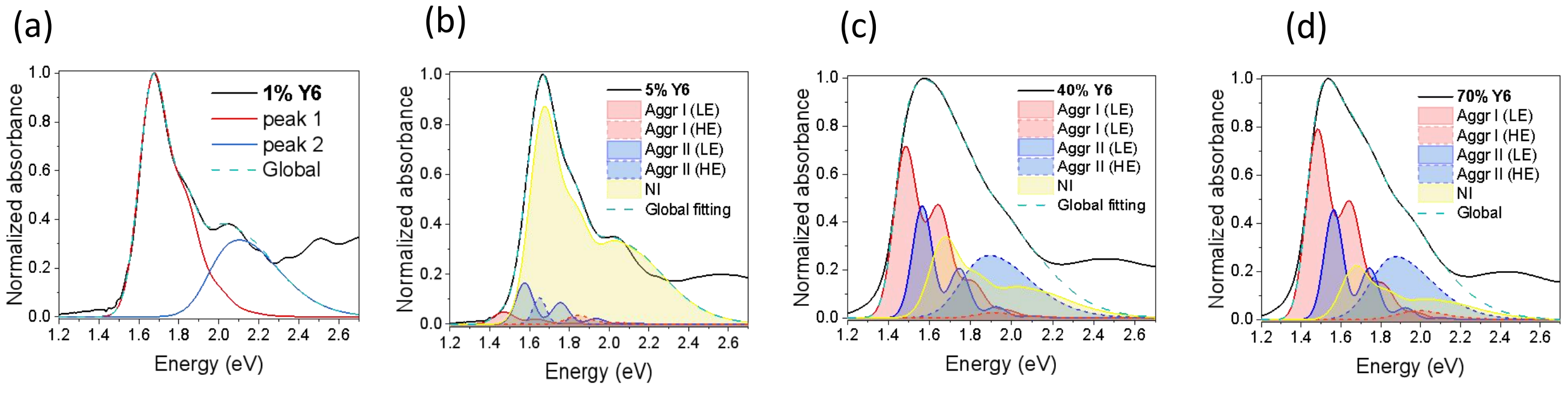

**Figure S2.** FCWD fitting result for (a) 1 wt%; (b) 5 wt% ; (c) 40 wt% and (d) 70 wt% Y6:PVK thin film

Based on the FCWD results, we therefore used the PL spectrum and PLQY at 1% or 2 wt% to represent monomer donors, and those at 10 wt% to represent aggregate II donors in parameterization of $R_0$. All parameters required for the FRET model were thus determined. The FRET efficiency of each pathway at all loading ratios was subsequently calculated using the software FRETCalc [Ref 43 in main text].

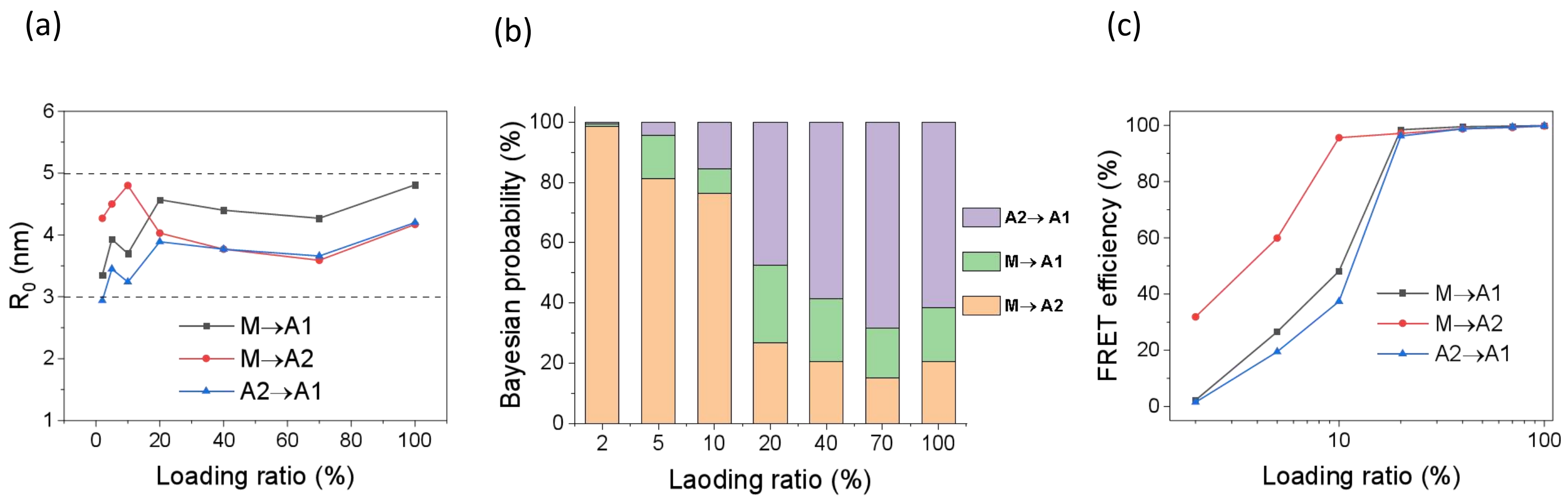

**Figure S3.** The variation trend of (a) Förster radius $R_0$; (b) corresponding Bayesian probability against loading ratio and (c) overlapping area of normalized extinction coefficient of aggregate I and II in pure Y6 as well as PL spectra of monomer and aggregate II PL

As mentioned in the main text, the overall FRET efficiency $E_{FRET}$ should consider the ensemble-weighted average of the FRET efficiency of all possible channels from donor-centric perspective. The expression of $E_{FRET}$ is given by **Eq. S6**.

$$E_{FRET} = E^{eff}_{M\rightarrow A2} + E^{eff}_{M\rightarrow A1} + E^{eff}_{A2\rightarrow A1} \quad \text{Eq S6.}$$

Here, $E^{eff}_{i\rightarrow j}$ is the expectation of FRET efficiency of a given channel $i \rightarrow j$, where the subscripts M, A1, and A2 refer to the energy transfer pathway.

In detail, considering the contribution from ensemble-weighted average of the FRET efficiency of all possible channels. The expression of $E_{FRET}$ can be written in **Eq. S7**.

$$E_{FRET} = \phi_M E_M + \phi_{A2} E_{A2} + \phi_{A1} E_{A1} \qquad \text{Eq. S7}$$

Subscripts M, A1, and A2 refer to the monomer and two aggregates. $E_M$ is the ensemble-averaged FRET efficiency for a single downhill transfer from a monomer donor to its nearest acceptor, and $\phi_M$ is the corresponding weight, namely the probability that an exciton initially resides on a monomer, as obtained from **Figure 2(e)**. $E_{A2}$ and $\phi_{A2}$ are defined analogously. As aggregate I is the lowest-energy acceptor species, it does not participate in downhill energy transfer, so $E_{A1}$= 0. As noted above, a monomer donor can transfer via two pathways. Therefore, **Eq. S7** can be further expanded as depicted in **Eq. S8**, where the corresponding weight of each pathway is the product of monomer exciton number density and the conditional probability that the relevant acceptor (A1 or A2) is the nearest neighbor.

$$E_{FRET} = \phi_M \left( \frac{\phi_{A2}}{\phi_{A2}+\phi_{A1}} E_{M\to A2} + \frac{\phi_{A1}}{\phi_{A2}+\phi_{A1}} E_{M\to A1} \right) + \phi_{A2} E_{A2\to A1} \qquad \text{Eq. S8}$$

The above expression shows that microscopic FRET of each pathway and the macroscopic ensemble averaged observable are linked by the initial exciton distribution, which arises from the partitioning of exciton among the different species and is directly governed by their number densities and aggregation behavior. The weight $w_{i\to j}$ of successful downhill energy transfer from the corresponding donor $i$ to acceptor $j$ constitutes the likelihood, taking the form of the weight preceding each efficiency term in **Eq. S8**. Therefore, the expression of $E_{FRET}$ can be as followed:

$$E_{FRET} = w_{M\to A2} E_{M\to A2} + w_{M\to A1} E_{M\to A1} + w_{A2\to A1} E_{A2\to A1} \qquad \text{Eq. S9}$$

The coefficient before each efficiency term is the weight that represents the impact from aggregation, so the product itself combines both aggregation effect and contribution of each term, which is the expectation of effective FRET efficiency, namely the expected efficiency of successful downhill transfer for a given channel. The corresponding expression is:

$$E_{i\to j}^{eff} = w_{i\to j}\, E_{i\to j} \qquad \text{Eq. S10}$$

The aggregation dependence of $R_0$ and the energy transfer efficiency of each channel is shown in **Figure S3(a)**. For all three energy transfer channels, $R_0$ varies only slightly, remaining consistently in the range of 3 to 5 nm. This is comparable to values reported for conventional organic systems [46-50, maintext], validating the credibility of our physical modeling. For the three transfer channels, the efficiencies behave as described in the main text: they rise rapidly at low concentrations and plateau near the aggregation threshold. Notably, at low concentrations, the monomer to aggregate II efficiency is substantially higher, further confirming that FRET occurs primarily between the two high-energy species in this regime.

At high concentrations, however, the synergistic interplay among multiple channels comes into play, and the efficiency of each individual channel alone is insufficient to resolve the underlying microscopic mechanism. Therefore, to more thoroughly assess the contribution of each channel to $E_{FRET}$, particularly at high concentrations, we applied Bayesian analysis to dissect the microscopic role of each channel throughout the aggregation progression. The Bayesian probability for FRET from species $i$ to species $j$ is given by **Eq. S11**.

$$P(i \to j \mid FRET) = \frac{w_{i\to j}\, E_{i\to j}}{E_{FRET}} \qquad \text{Eq. S11}$$

Here, the FRET efficiency $E_{i\to j}$ of each channel serves as the prior in the Bayesian framework. The probability $w_{i\to j}$ of successful downhill energy transfer from the corresponding donor constitutes the likelihood, taking the form of the weight preceding each efficiency term in **Eq. S9**. $E_{FRET}$ then acts as the evidence to calculate the Bayesian probability. The results are summarized in **Figure S3(b)**.

Consistent with the conclusions above, below 10% loading ratio, the Bayesian probability for FRET from monomer to aggregate II exceeds or approaches 80%, confirming that at low concentrations, FRET proceeds predominantly via the monomer to aggregate II channel. At high concentrations, the Bayesian probability for FRET from aggregate II to aggregate I dominates, whereas the probabilities for monomer-donor channels are substantially lower. This indicates that while FRET proceeds from both monomer and aggregate II to aggregate I overall, the aggregate II to aggregate I channel is the primary pathway. This finding is consistent with the FCWD results in the main text: at high concentrations, the monomer number density steadily decreases while the aggregate number density continues to rise, rendering aggregate II to aggregate I the dominant FRET contributor under these conditions.

**Figure S3(c)** presents the normalized extinction coefficients of the two aggregates in neat Y6 films and their spectral overlap with the PL spectra of the two donors. Aggregate I shows considerable overlap with both donor emissions, indicating that excitons in neat Y6 have a high probability of ultimately reaching aggregate I via FRET.

The collection efficiency of excitons that ultimately reach the Y-type aggregates, $E_{collection}$, is defined as below:

$$E_{collection} = \phi_{A1} + \frac{\phi_M \phi_{A1}}{\phi_{A2}+\phi_{A1}} E_{M\to A1} + \phi_{A2} E_{A2\to A1} \qquad \text{Eq. S12}$$

## Section 3. TAS

**Figures S4** and **S5** present the TA spectra, including the visible and near-infrared (NIR) ranges, respectively, at various delay times.

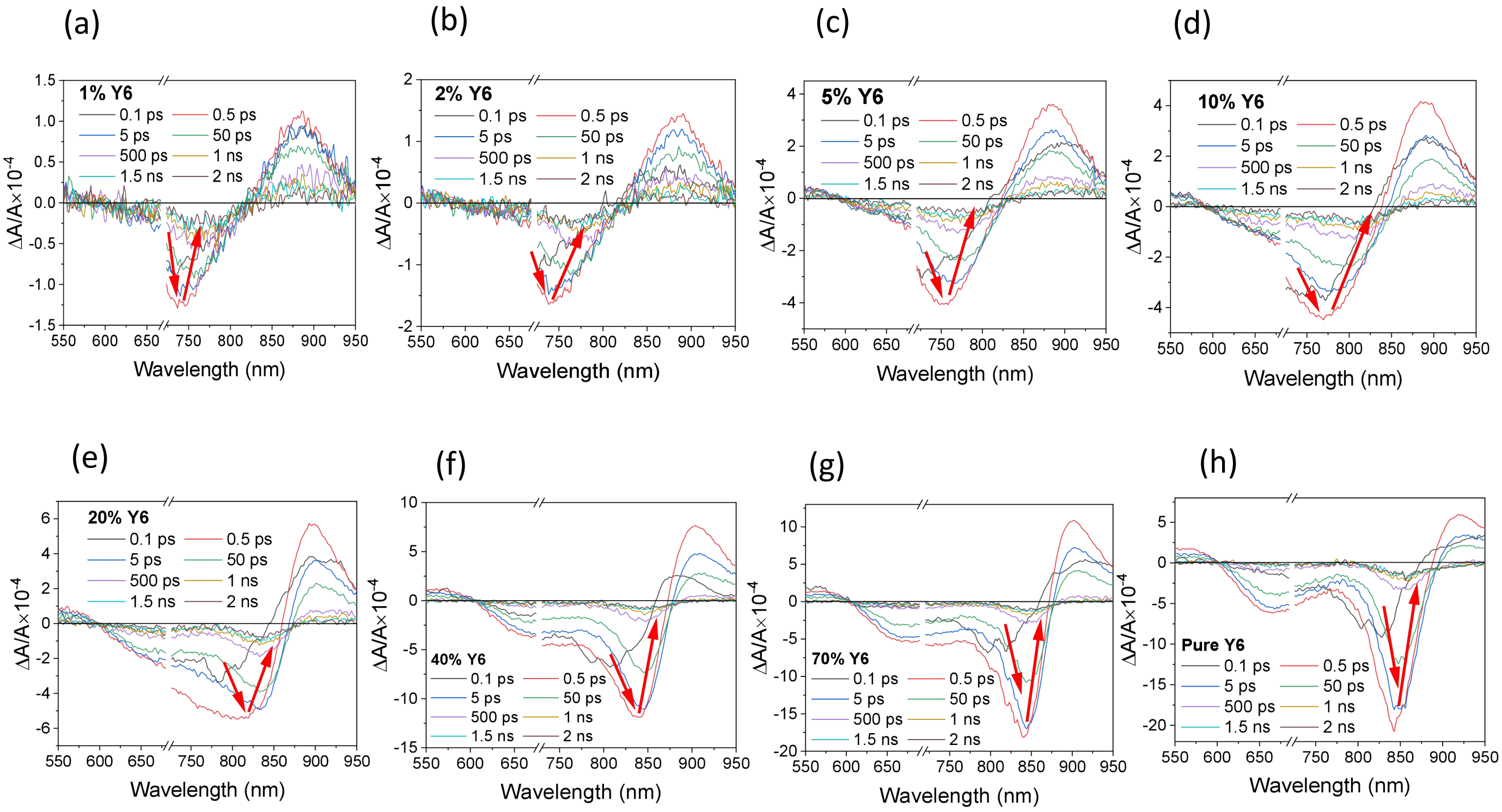

**Figure S4.** TA spectra with different delay times of (a) 1 wt%; (b) 2 wt%; (c) 5 wt%; (d) 10 wt%; (e) 20 wt%; (f) 40 wt%; (g) 70 wt% and (h) 100 wt% Y6 in PVK thin film

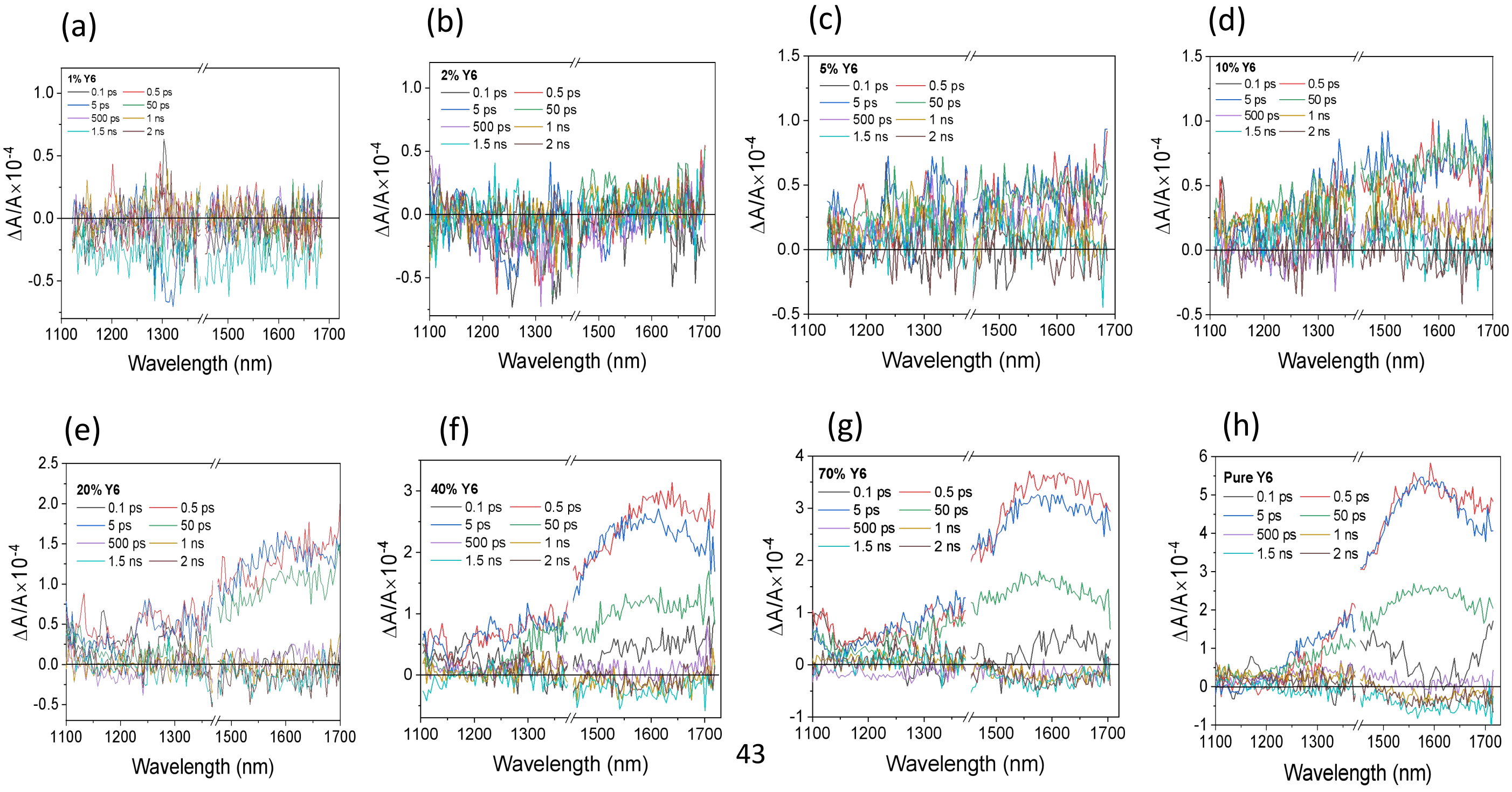

**Figure S5.** NIR TA spectra with different delay times of (a) 1 wt%; (b) 2 wt%; (c) 5 wt%; (d) 10 wt%; (e) 20 wt%; (f) 40 wt%; (g) 70 wt% and (h) 100 wt% Y6 in PVK thin film

The signal intensity in both spectral regions reaches its maximum at a delay time of 0.5 ps. Beyond this delay time, no new spectral features emerge, and only a monotonic decay in intensity is observed. Therefore, the TA profile recorded at 0.5 ps was selected for detailed analysis of each sample.

**Figures S6(a)** and **(b)** reveal a distinct spectral shift within the first 10 ps for pure Y6, which is absent in the 2 wt% sample. This difference can be attributed to efficient FRET occurring between energetically distinct species in the neat Y6 film, which is a process suppressed in diluted samples due to the lack of lower energy acceptor states.

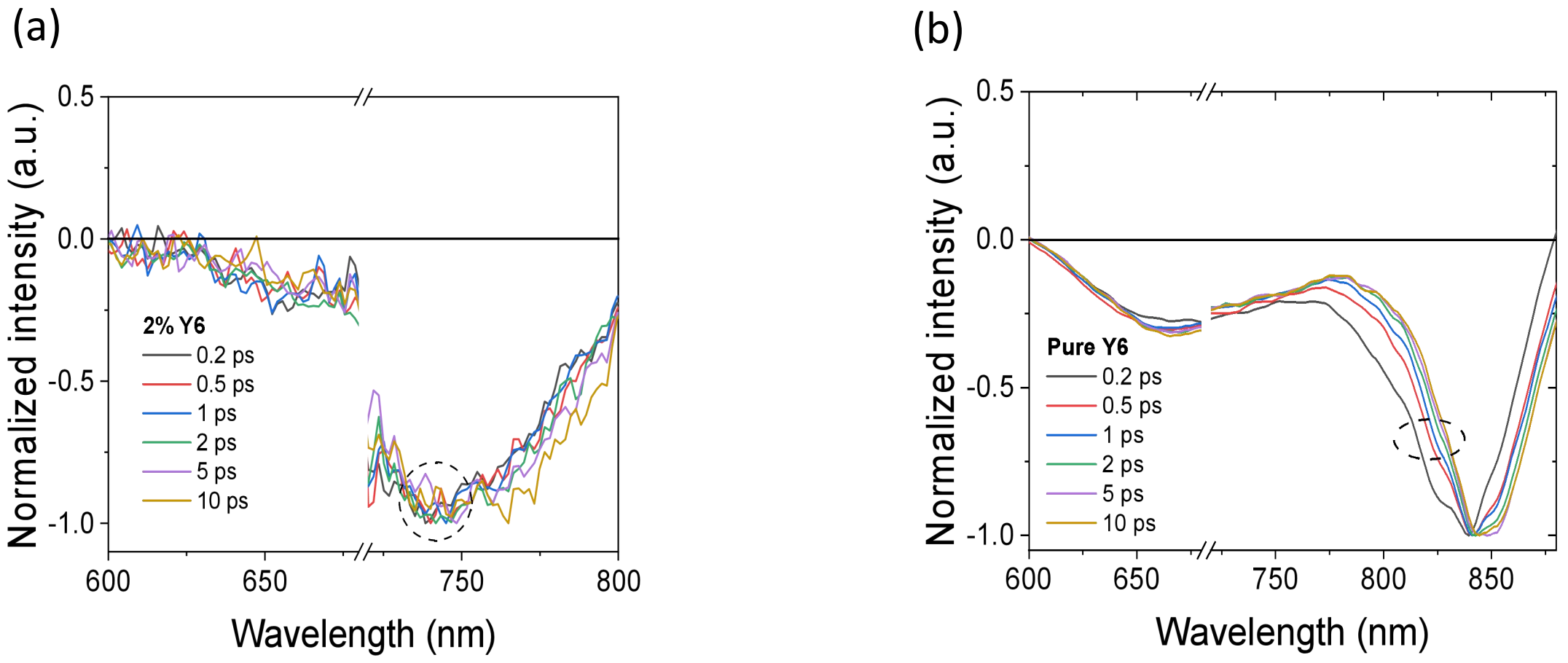

**Figure S6.** TA spectra probed at small delay time of (a) 2% and (b) pure Y6

## Section 4. Theoretical simulation

To obtain a suitable configuration for theoretical studies, we performed DFT optimization on the crystal structure of Y6 with simplified alkyl side chains (**Figure S7(a)**). From the optimized structure, we extracted three representative dimer configurations, named after their shapes: dimer S, dimer C, and dimer Y (**Figures S7(b)–S7(d)**). We then characterized the geometric parameters and interaction energies of these three dimer configurations.

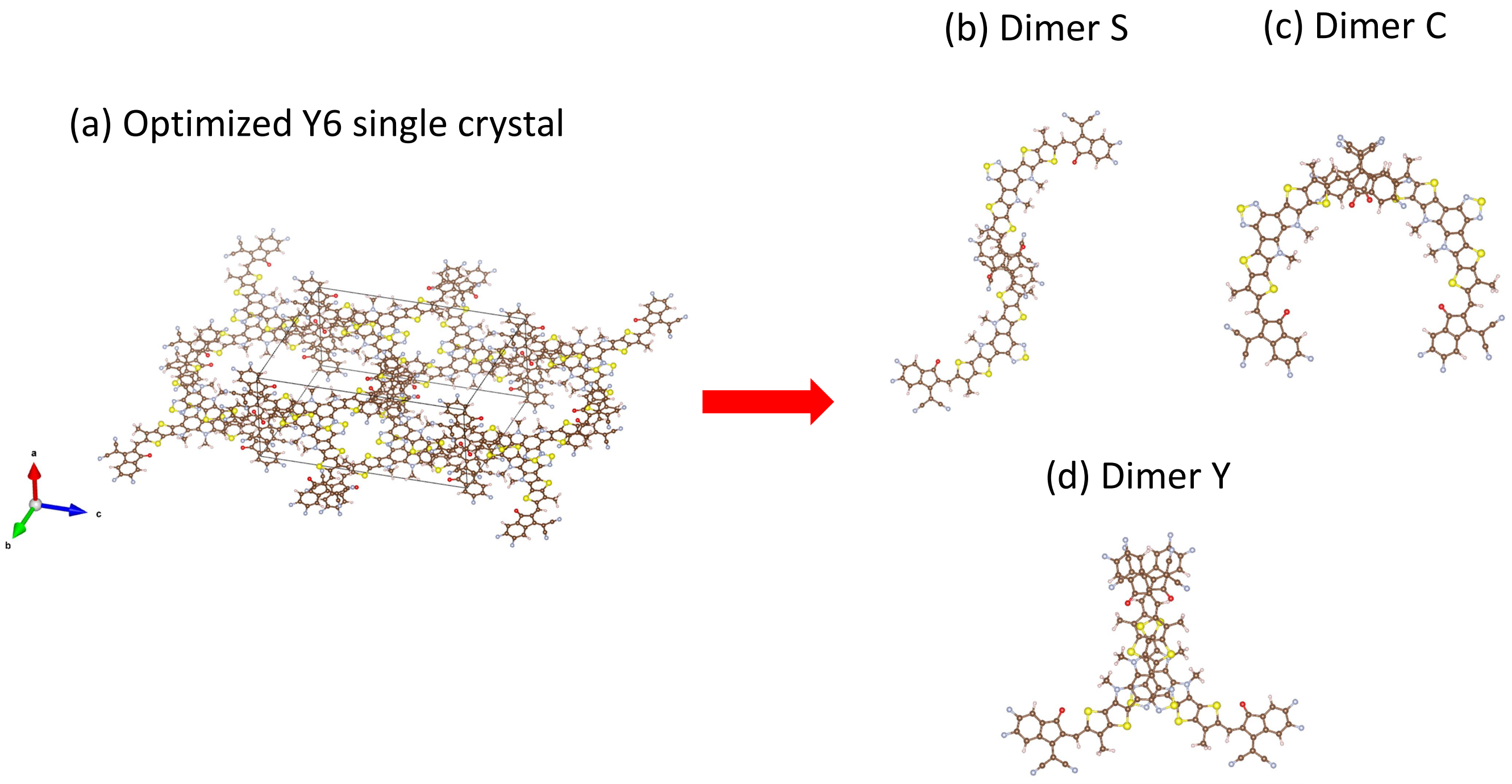

**Figure S7**. **Determination of the geometry configuration of aggregates I and II through DFT calculation.** (a) Optimized Y6 single crystal by DFT method; three typical dimer configurations extracted from optimized single crystal named by their shape as: (b) Dimer S ; (c) Dimer C and (d) Dimer Y

As a result, the centroid distance for S, C, and Y dimers was 1.89 nm, 1.29 nm, and 0.80 nm, respectively. While the interaction energy for them was -70.78 kJ/mol, -67.84 kJ/mol, and -144.31 kJ/mol, respectively, as shown in **Table S2**.

**Table S2.** Property summary of each species

| Component | Centroid distance (nm) | $E_{int}$ (kJ/mol) | CT character |
|---|---|---|---|
| Y6 Monomer | NA | NA | Weak |
| Dimer S | 1.89 | -70.78 | Weak |
| Dimer C | 1.29 | -67.84 | Intermediate |
| Dimer Y | 0.80 | -144.31 | Strong |

To further elucidate the excited state properties of the major species in Y6 films for validating the above assignments and for guiding the analysis of their CT characteristics, we performed TDDFT calculations for the first five excited states of the three primary species as mentioned above.

## Section 5. Methods

### 5.1 Materials and sample preparation

The non-fullerene acceptor Y6 was purchased from Tin Hang Technology Limited, and the polymer PVK was purchased from Sigma-Aldrich Corporation. Y6 and PVK were dissolved in chloroform individually to form a solution with both concentrations of 20 mg/ml. The quartz substrates were cleaned by sequential Decon 90, deionized water and alcohol. The cleaned substrates were then dried using a high-purity nitrogen gun and placed in a drying oven for 10 minutes before spin coating. For the diluted Y6 films, a different loading ratio (in wt%, 1%, 2%, 5%, 10%, 20%, 40%, and 70%) of Y6 was added to the PVK polymer solution to prepare the required concentration. Pure Y6 and diluted solution were spin-coated on quartz to form thin films of approximately 100 nm thickness inside a nitrogen-filled glovebox without annealing. The spin-coated thin films were fabricated for other characterizations.

### 5.2 UV-vis absorption measurement

Thin films with different loading ratios spin-coated on quartz substrates were measured using a Perkin-Elmer Lambda 1050 + UV/Vis/NIR spectrophotometer from 400 nm to 1000 nm.

### 5.3 PL & TRPL measurement

The PL and TRPL measurement for each sample was carried out with FLS1000 Photoluminescence Spectrometer from Edinburgh Instrument Corporation. The excitation wavelength in the TRPL measurement was 670 nm based on the absorption edge in the absorption measurement.

### 5.4 PLQY measurement

The PLQY measurement was achieved by using a home-made setup as shown in Figure S8 based on the method as follows. A laser with a wavelength of 673 nm served as the light source, while the QE Pro spectrometer from Ocean Optics was used to record the spectrum profiles.

Firstly, PLQY can be simply expressed by the number of photons emitted over the photons absorbed or defined as **Eq. 2** in the main text. In order to obtain the PLQY by experiment, there were three experiments named A, B and C to conduct according to the method proposed by de Mello et al [6] in 1997.

In experiment A, the laser beam traveled into the integrating sphere with no sample in the sphere. Thus, a spectrum of the laser can be obtained, as shown in **Figure S9 (a)**. This could provide information on the total intensity of unabsorbed light in the unit of count, which was the range between the blue lines. This range of laser profile was denoted as La.

In experiment B, the laser beam traveled into the integrating sphere without directly hitting the sample, as illustrated in **Figure S9 (b)**. A small portion of the light that was refracted by the refractive coating would be reabsorbed by the material, causing the material to emit photons. The reabsorbed light may cause a drop in the total intensity of the unabsorbed light. Thus, the spectrum obtained in the experiment could be divided into two parts, which were the laser profile and the PL profile. The laser profile and the PL profile obtained in the experiment were denoted as Lb and Pb, respectively.

Finally, in experiment C, the laser was directly hit on the sample as shown in **Figure S9 (c)**. Thus, a large portion of the light would be absorbed by the sample, and light would be emitted by the sample. The resulting spectrum of experiment C could also be divided into the laser profile and PL profile, just like experiment B, but with a much higher PL peak for the experiment. Similarly, the laser profile and the PL profile obtained in experiment C was denoted as Lc and Pc, respectively.

Now that the wavelength distribution and the intensity of the unabsorbed light and emitted light could be known in three experiments, the results, however, cannot be directly used to calculate the PLQY as the unit was in count. Therefore, the count of the laser profile need to be multiplied by the count of PL profile with the photons per count at different wavelengths obtained from the calibration so that the number of photons at different wavelengths could be obtained. Then integration area under the laser profiles and PL profile were needed to get the total number of photons absorbed and emitted in each experiment, namely the five constants La, Lb, Lc, Pb and Pc. This method accounts for the reabsorption effect, excluding photons emitted from refracted light, thereby providing a more accurate assessment of the material's intrinsic PLQY. Once the five parameters were obtained, one may calculate the PLQY with following equation:

$$\text{PLQY} = \frac{Pc-(1-A)\times Pb}{La\times A} \times 100\% \qquad \text{Eq. S13}$$

The A is the absorption coefficient defined as:

$$A = 1 - \frac{Lc}{Lb} \quad \text{Eq. S14}$$

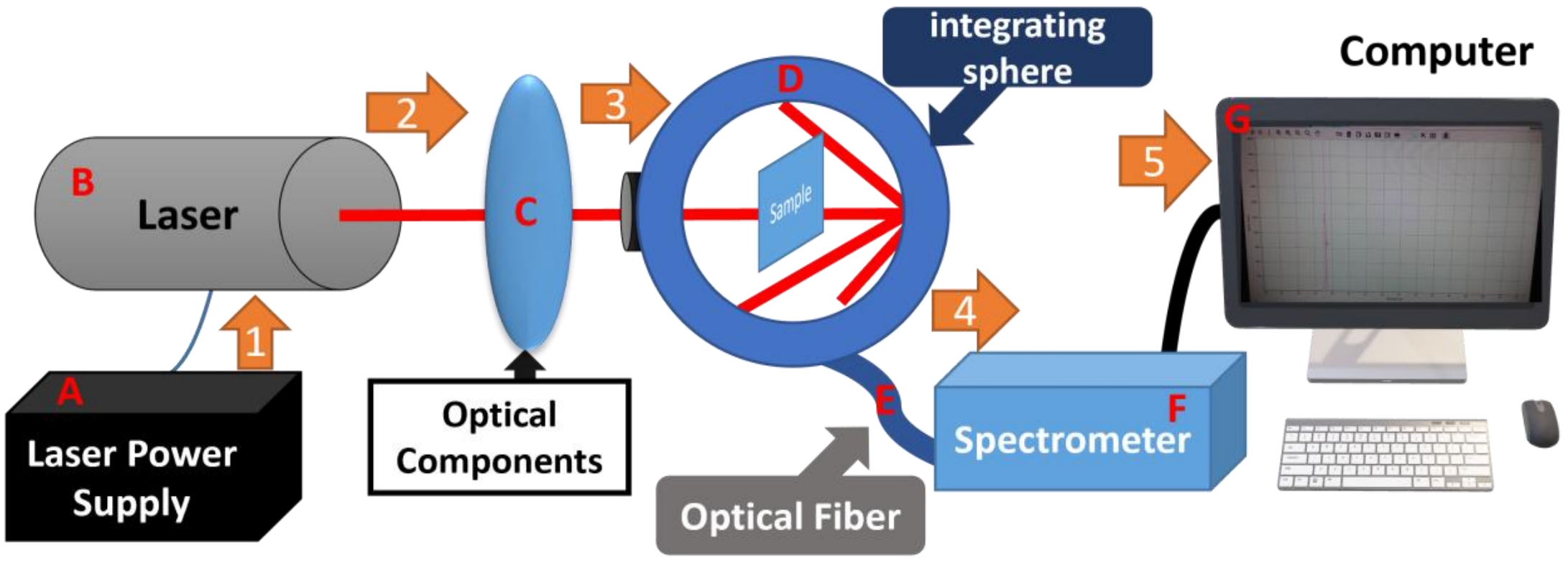

(a)

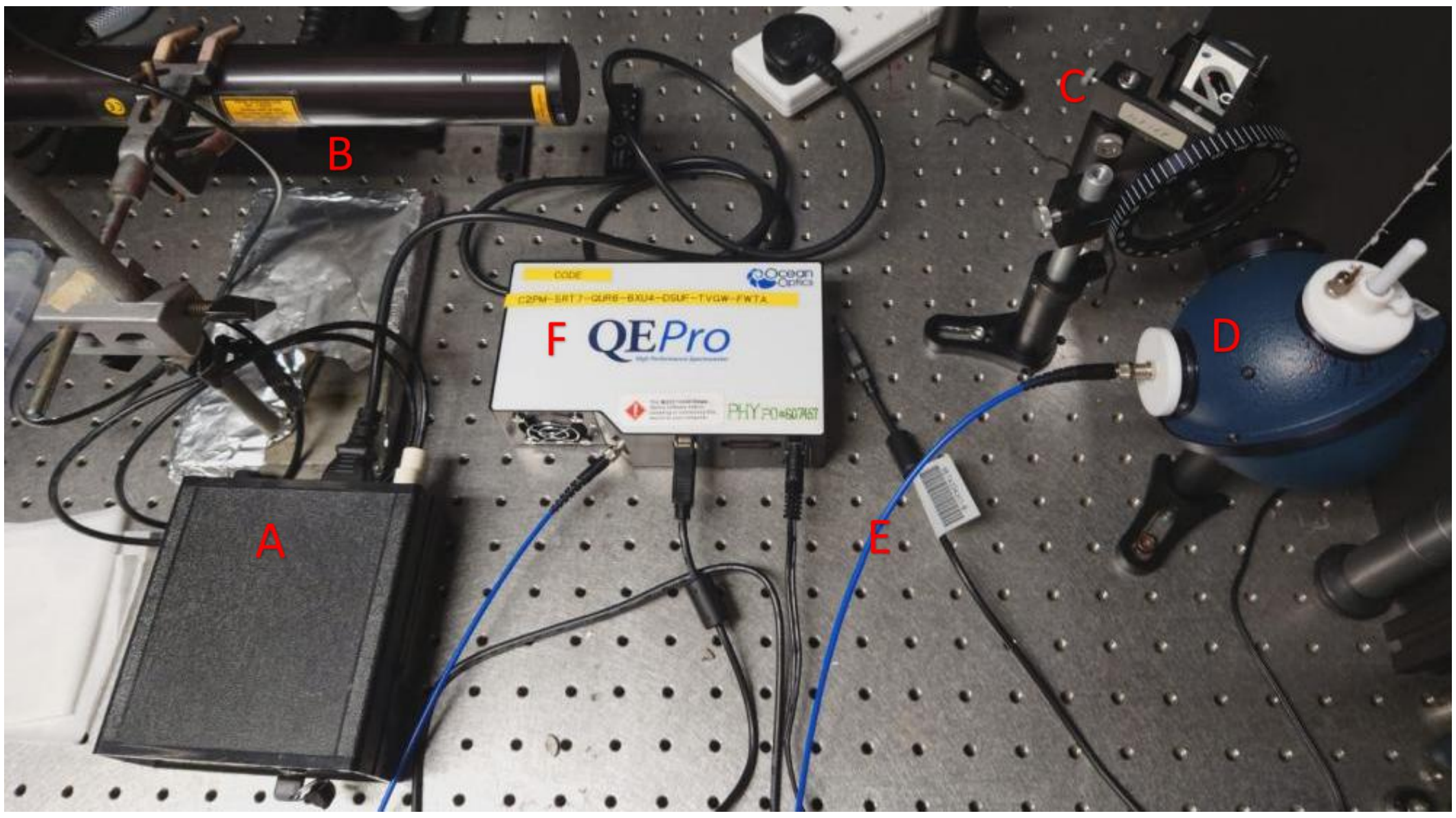

(b)

**Figure S8.** (a) Schematic and (b) A photograph of the home-made set for PLQY measurement

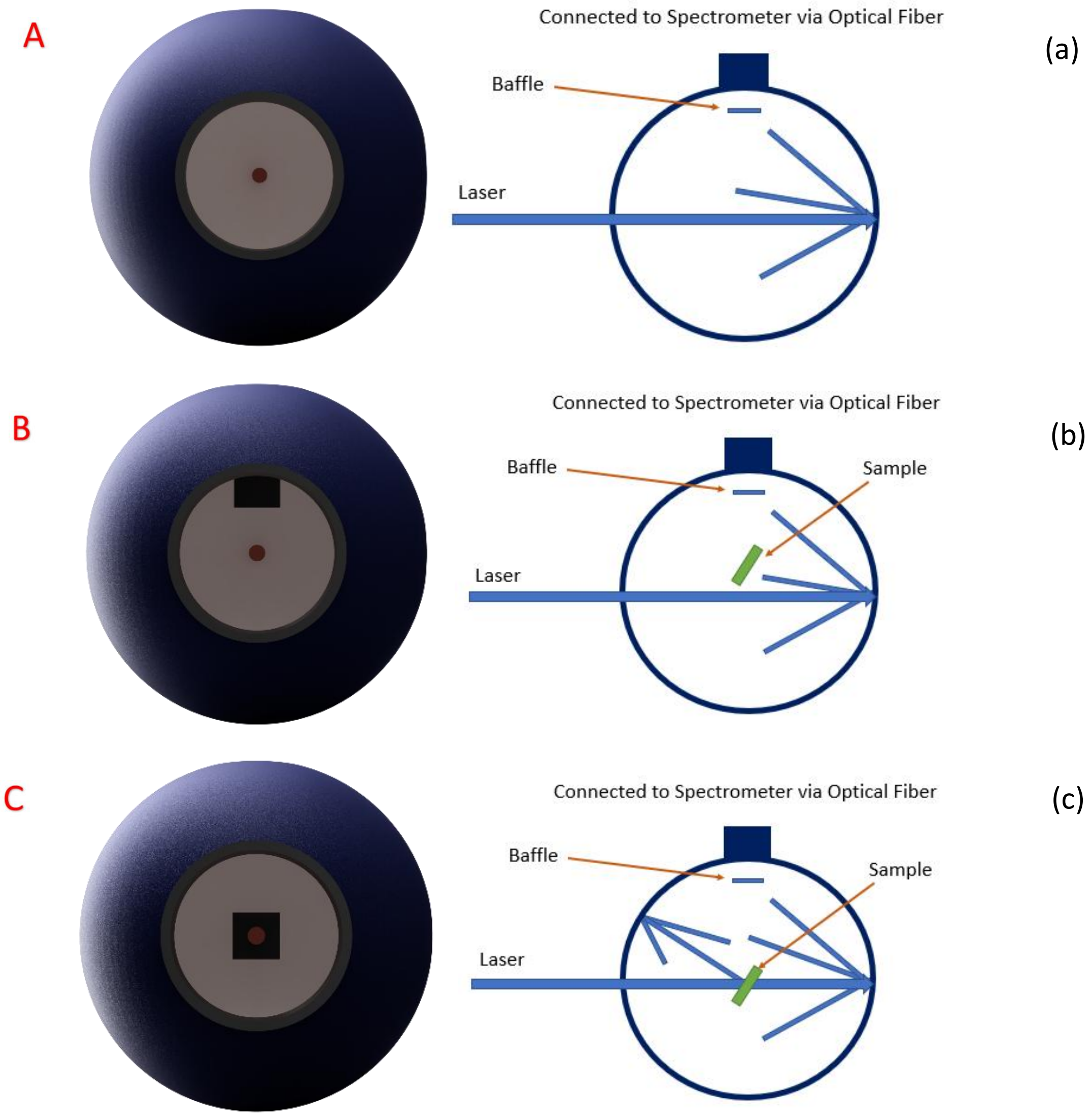

**Figure S9.** The configuration of sample in the three experiments for PLQY. Example of (a) no sample in the integrating sphere; (b) The laser beam is not directly hit on the sample and (c) the laser beam directly hit on the sample.

## 5.5 TAS measurement

TA measurements were performed using a Yb:KGW laser (Pharos, Light Conversion) with a fundamental output at around 1030 nm. The 700 nm pump pulses were generated by a second-harmonic noncollinear optical parametric amplifier (2HNOPA, Orpheus-N-2H, Light Conversion), with the output wavelength selected as required.

Probe pulses were produced via supercontinuum generation by focusing a small fraction of the 1030 nm fundamental into a sapphire plate for visible (VIS) detection or a yttrium aluminium garnet (YAG) plate for near-infrared (NIR) detection. The pump-probe time delay was controlled by a delay line. TA spectra were recorded on a commercial system (TA100, TIME TECH SPECTRA). Pulse-to-pulse spectral analysis was carried out at 10 kHz for VIS detection, achieving a signal-to-noise ratio better than $1 \times 10^{-5}$ after averaging 5000 pump-on and pump-off shots per data point. Unless otherwise specified, the pump fluence was set to about 2 μJ $cm^{-2}$. Samples were kept under a nitrogen atmosphere during measurements to prevent photodegradation.

### 5.6 Theoretical simulation

The first principle simulation was applied on Y6 single crystal at first to obtain the dimer and monomer structure in interest. The crystallographic data of Y6 single crystal could be checked in the Cambridge Crystallographic Data Centre (CCDC) by the series number as 2015912. The alkyl side chain of each Y6 molecule in the single crystal was replaced by methyl group for simplification. The geometry optimization of the simplified Y6 single crystal was achieved by CP2K 2024.2 package [7] using PBE functional with Grimme's DFT-D3(BJ) dispersion correction [8] and DZVP-MOLOPT-SR-GTH basis set. A 3 × 3 × 1 k point sampling was set using the Monkhorst-Pack scheme referring to early work [9]. The structure of three representative dimers and Y6 monomer were extracted from the optimized single crystal for following study.

The quantum chemical simulation was then applied on the dimers and monomer from optimized structure using ORCA 6.0.1 package [10]. The interaction energy of each dimer was calculated under the level of ωB97M-V/def2-TZVP with geometry counterpoise (gCP) correction to reduce the impact of basis set superposition error (BSSE) issue. The TDDFT method was applied to calculate five excited states for monomer and all dimers under the level of PBE0-D3(BJ)/def2-SVP. A Solvation Model Density (SMD) solvent model was applied in all quantum chemical calculation with diphenylether (ε=3.73) solvent to simulate the dielectric environment in the solid thin film. The hole-electron wavefunction analysis method was used based on the TDDFT results to study the CT properties of these species more quantitatively.

The visualization was achieved with the assistance of VESTA [11] and Multiwfn program [12,13]. Besides visualization, the preparation of input files for the calculation and the wavefunction analysis were achieved by Multiwfn program as well.

## Section 6. References

[1]. Ouyang, Y., Wang, R., Wang, X., Xiao, M. & Zhang, C. Ultrafast energy transfer beyond the Förster approximation in organic photovoltaic blends with non-fullerene acceptors. *Science Advances* **11**, eadr5973 (2025).

[2]. D'Angelo, P. *et al.* Electrical transport properties characterization of PVK (poly N-vinylcarbazole) for electroluminescent devices applications. *Solid-state electronics* **51**, 123-129 (2007).

[3]. He, X. *et al.* Selenium substitution for dielectric constant improvement and hole-transfer acceleration in non-fullerene organic solar cells. *Nature Communications* **15**, 2103 (2024).

[4]. Mallet, P., Guérin, C.-A. & Sentenac, A. Maxwell-Garnett mixing rule in the presence of multiple scattering: Derivation and accuracy. *Physical Review B—Condensed Matter and Materials Physics* **72**, 014205 (2005).

[5]. Jamnik, J., Kalnin, J., Kotomin, E. & Maier, J. Generalised Maxwell-Garnett equation: application to electrical and chemical transport. *Physical Chemistry Chemical Physics* **8**, 1310-1314 (2006).

[6]. de Mello, J. C., Wittmann, H. F. & Friend, R. H. An improved experimental determination of external photoluminescence quantum efficiency. *Advanced*

*materials* **9**, 230-232 (1997).

[7]. Kühne, T. D. *et al.* CP2K: An electronic structure and molecular dynamics software package-Quickstep: Efficient and accurate electronic structure calculations. *The Journal of Chemical Physics* **152** (2020).

[8]. Grimme, S., Antony, J., Ehrlich, S. & Krieg, H. A consistent and accurate ab initio parametrization of density functional dispersion correction (DFT-D) for the 94 elements H-Pu. *The Journal of chemical physics* **132** (2010).

[9]. Zhang, G. et al. Delocalization of exciton and electron wavefunction in non-fullerene acceptor molecules enables efficient organic solar cells. Nature communications 11, 3943 (2020).

[10]. Neese, F. Software update: The ORCA program system—Version 5.0. *Wiley Interdisciplinary Reviews: Computational Molecular Science* **12**, e1606 (2022).

[11]. Momma, K. & Izumi, F. VESTA 3 for three-dimensional visualization of crystal, volumetric and morphology data. *Applied Crystallography* **44**, 1272-1276 (2011).

[12]. Lu, T. & Chen, F. Multiwfn: A multifunctional wavefunction analyzer. *Journal of computational chemistry* **33**, 580-592 (2012).

[13]. Lu, T. A comprehensive electron wavefunction analysis toolbox for chemists, Multiwfn. *The Journal of chemical physics* **161** (2024).